\documentclass[journal]{IEEEtran}
  \usepackage[ruled,linesnumbered]{algorithm2e}
  \usepackage{amsmath}
  \usepackage{booktabs}
  \usepackage{color,soul}
  \soulregister\cite7 % 针对\cite命令
  \soulregister\citep7 % 针对\citep命令
  \soulregister\citet7 % 针对\citet命令
  \soulregister\ref7 % 针对\ref命令
  \soulregister\pageref7 % 针对\pageref命令
  \soulregister\textbf7
  \soulregister\section7
  \usepackage{diagbox}
  \usepackage{makecell}
  \usepackage{multirow,multicol}
  \usepackage{soul}
  \usepackage{textcomp}
  \usepackage{subfig}
  \usepackage{tabularx}
  \usepackage{threeparttable}
  \usepackage{footnote}
  \usepackage{caption}
  \usepackage{graphicx}
\ifCLASSINFOpdf
\else
\fi

\begin{document}
\title{iTV: Inferring Traffic Violation-Prone Locations with Vehicle Trajectories and Road Environment Data}

\author{Zhihan~Jiang,
        Longbiao~Chen,~\IEEEmembership{Member,~IEEE},
        Binbin~Zhou,
        Jinchun~Huang,
        Tianqi Xie,\\
        Xiaoliang Fan,~\IEEEmembership{Member,~IEEE},
        Cheng Wang,~\IEEEmembership{Member,~IEEE}

\thanks{Manuscript received March 14, 2020; revised June 11, 2020; accepted July 21, 2020. Date of publication ...; date of current version July 27, 2020. (Corresponding author: Longbiao Chen)}

\thanks{Z. Jiang, L. Chen, J. Huang, T. Xie, X. Fan and C. Wang are with Fujian Key Laboratory of Sensing and Computing for Smart Cities, School of Informatics, Xiamen University, Xiamen 361005, China (e-mail: zhihanjiang@stu.xmu.edu.cn, longbiaochen@xmu.edu.cn).}

\thanks{B. Zhou is with Zhejiang University, Hangzhou 310000, China (e-mail: bbzhou@zju.edu.cn).}% <-this % stops a space
\thanks{Digital Object Identifier 10.1109/JSYST.2020.3012743}
\thanks{\textbf{Copyright: 0000--0000/00\$00.00~\copyright~2020 IEEE}}
}
% \vspace{-1cm}
% \thanks{\IEEEpubid{0000--0000/00\$00.00~\copyright~2020 IEEE}}


\markboth{IEEE SYSTEM JOURNAL,~Vol.~xx, No.~xx, xx~20xx}%
{Z. Jiang \MakeLowercase{\textit{et al.}}: Bare Demo of IEEEtran.cls for IEEE Journals}

% \IEEEpubid{0000--0000/00\$00.00~\copyright~2020 IEEE}


\maketitle

\begin{abstract}
  Traffic violations like illegal parking, illegal turning, and speeding have become one of the greatest challenges in urban transportation systems, bringing potential risks of traffic congestions, vehicle accidents, and parking difficulties. To maximize the utility and effectiveness of the traffic enforcement strategies aiming at reducing traffic violations, it is essential for urban authorities to infer the traffic violation-prone locations in the city. Therefore, we propose a low-cost, comprehensive, and dynamic framework to infer traffic violation-prone locations in cities based on the large-scale vehicle trajectory data and road environment data. Firstly, we normalize the trajectory data by map matching algorithms and extract key driving behaviors, i.e., turning behaviors, parking behaviors, and speeds of vehicles. Secondly, we restore spatiotemporal contexts of driving behaviors to get corresponding traffic restrictions such as no parking, no turning, and speed restrictions. After matching the traffic restrictions with driving behaviors, we get the traffic violation distribution. Finally, we extract the spatiotemporal patterns of traffic violations, and build a visualization system to showcase the inferred traffic violation-prone locations. To evaluate the effectiveness of the proposed method, we conduct extensive studies on large-scale, real-world vehicle GPS trajectories collected from two Chinese cities, respectively. Evaluation results confirm that the proposed framework infers traffic violation-prone locations effectively and efficiently, providing comprehensive decision supports for traffic enforcement strategies.

\end{abstract}


% \fi
\IEEEpeerreviewmaketitle

\section{Introduction}

\label{sec:introduction}
  
  \IEEEPARstart{T}{raffic} violations, such as speeding and illegal parking, have become one of the greatest challenges in urban transportation systems, bringing potential risks of traffic congestions, vehicle accidents, and parking difficulties, etc. \cite{zhang2013risk,coric2013crowdsensing,moyano1997evaluation}. For example, in 2018, New York City witnessed 54,469 traffic violations and 44,508 traffic injuries across the city \cite{NewYorkCity_2019}. To reduce traffic violations, urban authorities have implemented various traffic enforcement strategies, such as deploying field enforcement officers in rush hours and installing automated monitoring cameras in road intersections \cite{TransportationAlternatives_2009}. Given the expensive human resource allocation and infrastructure investment, it is essential for urban authorities to identify the traffic violation-prone locations so as to deploy officers and install cameras under limited labor and non-labor resources.

  However, traditional strategies for traffic violation-prone location inference are highly dependent on historical traffic violation records and human experience, which are labor intensive, time consuming, and unable to adapt to rapidly developing cities. Therefore, a low-cost, comprehensive, and dynamic method is in great demand. Fortunately, with the popularization of GPS devices and map services like street view service, we can get crowd-sensed and large-scale vehicle trajectory data in cities, real panoramic street view pictures on roads, and related traffic restrictions such as speed restrictions. These rich trajectory data and road environment data provide us an unprecedented opportunity to explore traffic violation-prone locations.  

  In this work, we propose a low-cost, comprehensive and dynamic framework for inferring the traffic violation-prone locations in cities based on the crowd-sensed, large-scale  vehicle trajectory data and road environment data fusion, so that we can provide some insights for the traffic management department about traffic violation-prone locations to help optimize the utility and effectiveness of the traffic enforcement strategies.

  Firstly, we normalize the trajectory data by mapping the vehicle trajectories onto the road network and get the driving behaviors. Secondly, we model driver perspectives to match driving behaviors to corresponding road segments and get the spatiotemporal contexts of driving behaviors. Using the spatiotemporal context, we detect traffic signs to identify no-turning road intersections and no-parking road segments. We can also get speed restrictions on roads from real-time navigation service providers. After matching the traffic restriction information with driving behaviors, we extract three types of traffic violations, i.e., illegal turning, illegal parking and speeding, and extract the spatiotemporal patterns of traffic violations to infer the traffic violation-prone locations. Finally, we build a traffic violation-prone locations inference system and evaluate the proposed method using large-scale, real-world datasets from two cities in China, Chengdu and Xiamen.

    \begin{figure}[!t]
        \vspace{-0.5cm}
        \centering
        \subfloat[A vehicle trajectory]{
                \centering
                \includegraphics[width=.12\textwidth]{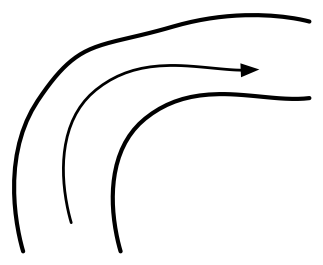}
        \label{fig:extract_directly_1}
        }\quad\quad
        \subfloat[A sparse trajectory]{    
                \centering
                \includegraphics[width=.12\textwidth]{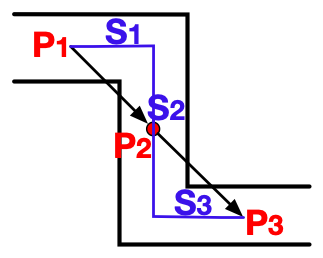}
            \label{fig:extract_directly_2}
        }\quad\quad
        \subfloat[A dense trajectory]{ 
                \centering
                \includegraphics[width=.12\textwidth]{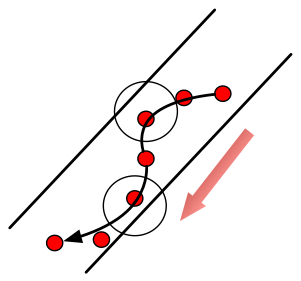}
            \label{fig:extract_directly_3}
        }
        \caption{Some misunderstanding examples of turning behaviors.}
        \label{fig:extract_directly}
    \end{figure}

    % \begin{figure}
    %     \centering
    %     \includegraphics[width=.4\textwidth]{pano_example.png}
    %     \caption{An example of street view picture.}
    %     \label{fig:pano_example}
    % \end{figure}

  In designing the framework, there are several research issues to be addressed: 
  % Ordered List -------------------------------------
  \begin{enumerate}
    \item \textbf{It is non-trivial to extract turning behaviors from vehicle GPS trace data}. When only trajectory data is used, it is easy to misunderstand some driving behaviors. For example, as shown in \figurename~\ref{fig:extract_directly_1}, if we observe the trajectory separately, we may consider that it is a turning behavior in an intersection. However, if we locate the trajectory into the corresponding road network, we can find that it is not a turning behavior since it is caused by the curvy road rather than the driver's decision in the intersection. Moreover, some sparse GPS points will also mislead us. As shown in \figurename~\ref{fig:extract_directly_2}, the trajectory $P_1\to P_2\to P_3$ implies us an straight line, while it should be $S_1\to S_2\to S_3$ actually when combined with the road network. Here, two turning behaviors can be extracted. Besides, if the GPS points are dense and the driver prefers changing lanes while driving, there will be several curves on the trajectory, as shown in the circled parts in \figurename~\ref{fig:extract_directly_3}. Then those curves will be mistaken for turning behaviors. Therefore, to extract turning behaviors accurately, we should take both GPS trajectories and road networks into consideration.

    \item \textbf{It is difficult to restore the spatiotemporal contexts of driving behaviors}. To identify whether a driving behavior is illegal or not, we need to restore the spatiotemporal context of driving behavior. For example, a driver makes an impermissible left turn in the intersection with a no-left-turn sign located in the driver's previous perspective; thus, this behavior can be identified as a traffic violation. However, with only vehicle GPS trajectories, it is difficult to get traffic restriction information for road segments where the trajectories are. Fortunately, the development of various map services enables us to get panoramic pictures of the surrounding environment of almost all streets in the city and real-time navigation service can offer us speed restriction information while driving. Those services enable us to restore spatiotemporal contexts of driving behaviors.

  \end{enumerate}

  In summary, the main contributions of this paper include:
  % Ordered List -------------------------------------
  \begin{enumerate}

    \item To the best of our knowledge, this is the first work using crowd-sensed and large-scale vehicle trajectory and road environment data to infer traffic violation-prone locations in cities. Such a low-cost, dynamic, comprehensive method can help urban authorities maximize the utility and effectiveness of the traffic enforcement strategies.

    \item We proposed a data-driven method to identify traffic violation-prone locations. Firstly, we normalize the GPS trajectories with map matching methods and get the driving behavior distribution. Secondly, we model driver perspectives based on regression to augment the spatiotemporal contexts of driving behaviors. Thirdly, we extract three types of traffic restrictions from those contexts, i.e., no-parking, no-turning, and speed restrictions. After matching those restrictions with driving behaviors, we get the traffic violation distribution and extract the spatiotemporal features of traffic violations. Finally, we build the traffic violation-prone location inference system.

    \item We evaluate the proposed method using large-scale, real-world datasets from two Chinese cities, i.e., Chengdu and Xiamen, including vehicle GPS trajectories, street view pictures, and speed restrictions. The experimental results are visualized in a traffic violation-prone location inference system.

  \end{enumerate}

  % The rest of this paper is organized as follows. We begin by reviewing related work in Section II. Then we introduce the definitions related to this work and present an overview of the proposed framework in Section III. In the following three sections, we detail the steps of driving behavior extraction, behavior context restoration, and traffic violation-prone location inference. And in section VII, we evaluate the framework based on real-world data from two cities. Finally, we conclude and introduce future work in the last section.

% ==========================================================
\section{Related Work}

  % In this section, we review the related work in three groups. The first group consists of different applications of GPS trajectory mining, the second group introduces researches in traffic sign detection, and the third group reviews the applications of the map-matching method.

  \subsection{GPS Trajectory Mining}

  In the literature, there have been many studies on mining various kinds of GPS trajectories for different application scenarios \cite{chen2016crowddeliver}. In a study on human mobility, Zheng et al. \cite{zheng_mining_2009} proposed a framework to mine interesting locations and travel sequences from human GPS trajectories, while Alvares et al. \cite{alvares_model_2007} proposed a model to enrich human GPS trajectories with semantic geographical meanings. On taxi operation study, Zhang et al. \cite{zhang_ibat:_2011} detected anomalous passenger delivery trips from taxi GPS traces, Chen et al. \cite{chen_bike_2015} explored citywide night bus planning issues leveraging taxi GPS traces, and Zhang et al. \cite{zhang_sensing_2013} identified taxi refueling behaviors from GPS trajectories to estimate citywide petrol consumption and analyze gas station efficiency. Chen et al. \cite{chen2016container} proposed a framework for container port performance measurement and comparison using ship GPS traces. As for driving violation identification, He et al. \cite{he2018detecting} proposed a framework to detect illegal parking events using shared bikes' trajectories. Lee et al. \cite{lee2009real} proposed a methodology for detecting illegal parking events in real-time by applying a novel picture projection that reduces the dimensionality of the data. Cerber et al. \cite{gerber1995vehicle} proposed a disclosed method of traffic control, which can detect and identify the vehicle speeding. In this work, the GPS trajectory data is one of the major resources, on which we can infer the traffic violation-prone locations.

  \subsection{Traffic Sign Detection}

  Traffic sign detection has been widely studied in the field of computer vision. Escalera et al. \cite{de1997road} used color thresholding and shape analysis to detect signs in pictures and classified signs with a neural network. Bahlmann et al. \cite{bahlmann2005system} detected signs using a set of Haar wavelet features obtained from AdaBoost training and classified signs using Bayesian generative modeling. Mogelmose et al. \cite{mogelmose2012vision} provided a survey of the traffic sign detection literature, detailing detection systems for traffic sign detection for driver assistance. Houben et al. \cite{houben2013detection} introduced a real-world benchmark data set from Germany for traffic sign detection and compared several detection approaches. In recent years, deep learning methods have shown superior performance for many tasks, such as object detection. Zhu et al. \cite{zhu2016traffic} proposed a framework of traffic sign detection and detection based on proposals by the guidance of a fully convolutional network. Zhu et al. \cite{Zhe_2016_CVPR} provided 100000 pictures containing 30000 traffic-sign instances and demonstrated how a robust end-to-end convolutional neural network (CNN) could simultaneously detect and classify traffic signs. Luo et al. \cite{luo2018traffic} proposed a data-driven system to detect traffic signs, which include both symbol-based and text-based signs, in video sequences captured by a camera mounted on a car. Huang et al. \cite{huang2018detection} introduced GAN into the Faster-RCNN framework and improved the performance of small object detection compared to Faster-RCNN. Yolov3 \cite{redmon2018yolov3} is a well-know state-of-art object detection system, which has achieved great performance both in accuracy and time. It has made great progress in small object detection and has strong generalization ability. In this work, we choose YOLOv3 as the backbone network to train the traffic sign detection model with transfer learning \cite{pan2009survey}. We first use the Chinese traffic-sign benchmark provided by \cite{Zhe_2016_CVPR} to pre-train a traffic sign detection model. Then we fine-tune the model based on a small dataset with less categories of traffic signs collected from the targeted city. In this way, the final model can be adapted to the task of interested categories of traffic signs and cities, and it can be transferred to new cities easier, since it only requires a small dataset to fine-tune the pre-trained model.

  \subsection{Map Matching Methods}

  Map matching is the procedure for mapping vehicle trajectories onto the correct road segments, which plays an important role in-vehicle navigation systems, routing engines, etc. \cite{newson2009hidden}. Ren et al. \cite{ren2009hidden} introduced the GPS-based wheelchair navigation based on a novel map matching algorithm. Bernstein et al. \cite{bernstein1996introduction} introduced the personal navigation assistants based on the map matching method. He et al. \cite{he2018detecting} used the map matching method to detect illegal parking events. White et al. introduced some map matching algorithms for personal navigation assistant \cite{white2000some}. Chen et al.\cite{chen2019trajcompressor} proposed an online map-matching-based trajectory compression framework running under the mobile environment. Vaughn et al. \cite{vaughn1996vehicle} described the GPS-map speed matching system for controlling the speed of the vehicle. The nearest matching method is the most basic map matching method, in which the latitude/longitude pairs are assigned to the nearest points or curves on the road network. However, this method is unable to take the contexts of GPS point sequences into account. Based on the nearest matching method, some environment factors, such as orientation information, connectivity constraints, and topologically feasible path through the road network, are incorporated to improve the performance of map matching \cite{white2000some, greenfeld2002matching}. However, these purely geometric methods are sensitive to measurement noise and sampling rate, while the Hidden Markov Model (HMM) addresses this issue by explicitly modeling the connectivity of the roads and considering many different path hypotheses simultaneously \cite{newson2009hidden, brakatsoulas2005map, dewandaru2007novel}. In this work, we normalize the trajectory data with a map matching method so as to extract driving behaviors, and since the trajectory data are noisy and nonuniform in real-world deployment, we choose to use the map matching algorithm based on HMM.

% =============================================================================
\section{Preliminary and Framework Overview}

  In this section, we introduce different kinds of traffic signs, define several terms used in this paper, and present the overview of the proposed framework.

  \subsection{Preliminary}

    \textbf{Road Network}: A road network can be defined as a graph $G=(I,R)$, where $I=\{i_1,i_2,...,i_n\}$ is a set of road intersections, and $R=\{r_1,r_2,...,r_m\}$ is a set of road segments.

   \textbf{GPS Point Trajectory}: A GPS point can be denoted as 4-tuples, $ p = (id,t,lat,lng) \}$, where $ id,t,lat,lng $ are the unique trajectory ID, time stamp, latitude, and longitude from GPS transmitters, respectively. A GPS point trajectory $traj_p$ can be defined as a time-ordered sequence $P_{s}=\{p_1\to p_2\to ...\to p_n\}$, where $n$ is the number of GPS points in the trajectory.

   \textbf{Road Segment Trajectory}: A road segment trajectory $traj_g$ can be denoted by 2-tuples, $traj_g=(id,R_{s})$, where $id$ is the unique trajectory ID, and $R_{s}$ is defined as a time-ordered sequence of road segments, $R_{s}=\{r_1\to r_2\to...\to r_m\}$, where $r_i\in R, 1\leq i\leq m$ and $m$ is the number of road segments in the trajectory. 

    %   \begin{figure}
    %   \centering
    %   \subfloat[]{
    %         \begin{minipage}[c]{.25\linewidth}
    %             \centering
    %             \includegraphics[width=\textwidth]{driving_behaviors_1.png}
    %         \end{minipage}
    %     \label{fig:driving_behaviors_1}
    %     }
    %     \quad
    %     \subfloat[]{    
    %         \begin{minipage}[c]{.25\linewidth}
    %             \centering
    %             \includegraphics[width=\textwidth]{driving_behaviors_2.png}
    %         \end{minipage}
    %         \label{fig:driving_behaviors_2}
    %     }
    %     \quad
    %     \subfloat[]{ 
    %         \begin{minipage}[c]{.25\linewidth}
    %             \centering
    %             \includegraphics[width=\textwidth]{driving_behaviors_3.png}
    %         \end{minipage}
    %         \label{fig:driving_behaviors_3}
    %     }
    %     \caption{An illustration of turning behaviors. (a) A turn-left behavior. (b) A turn-right behavior. (c) A u-turn behavior}
    %     \label{fig:driving_behaviors}
    % \end{figure}
    % \vspace{-1cm}
    \begin{figure}
      \centering
      \subfloat[]{
            \begin{minipage}[c]{.15\linewidth}
                \centering
                \includegraphics[width=\textwidth]{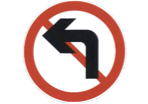}
            \end{minipage}
        \label{fig:traffic_signs_1}
        }
        \quad
        \subfloat[]{    
            \begin{minipage}[c]{.15\linewidth}
                \centering
                \includegraphics[width=\textwidth]{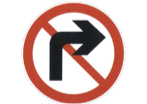}
            \end{minipage}
            \label{fig:traffic_signs_2}
        }
        \quad
        \subfloat[]{ 
            \begin{minipage}[c]{.15\linewidth}
                \centering
                \includegraphics[width=\textwidth]{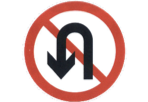}
            \end{minipage}
            \label{fig:traffic_signs_3}
        }
        \quad
        \subfloat[]{ 
            \begin{minipage}[c]{.15\linewidth}
                \centering
                \includegraphics[width=\textwidth]{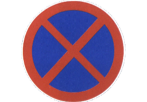}
            \end{minipage}
            \label{fig:traffic_signs_4}
        }
        \caption{Four most frequently used traffic sign types. (a) No-left-turn. (b) No-right-turn. (c) No-u-turn. (d) No-parking.}
        \label{fig:traffic_signs}
    \end{figure}

   \textbf{Turning Behavior}: The turning behavior can be defined as 8-tuples, $tn_i=(type,id,lat,lng,t,bb,ba,conf)$, where $id$, $lat$, $lng$, $t$, $bb$, $ba$, $conf$ are the unique trajectory ID, latitude, longitude, time of the turning behavior, $bb$ and $ba$ are the direction of the vehicle before and after the turning behavior, respectively, which are the clockwise angles from the North, $conf$ is the confidence value of the turning behavior and $type$ is the type of the turning behavior, $type\in\{left-turn,right-turn,u-turn\}$, corresponding to the three traffic signs shown in \figurename~\ref{fig:traffic_signs}. A set of turning behaviors is denoted by $TN=\{tn_1, tn_2,...,tn_m\}$, where $m$ is the number of turning behaviors.

   \textbf{Parking Behavior}: The parking behavior can be defined as 5-tuples, $pk_i=(id,lat,lng,st,et)$, where $id$, $lat$, $lng$, $st$, $et$ are the unique trajectory ID, longitude, latitude, start time and end time of the parking behavior, which is consistent with the no-parking sign shown in \figurename~\ref{fig:traffic_signs} (d). A set of parking behaviors is denoted by $PK= \{pk_1, pk_2,...,pk_m\}$, where $m$ is the number of parking behaviors.
    % \begin{figure}
    %     \centering
    %     \includegraphics[width=.2\textwidth]{Distance.png}
    %     \caption{Two distances.}
    %     \label{fig:dis}
    % \end{figure}

    % We also introduce the four most frequently used traffic signs in our study, i.e., no-left-turn, no-right-turn, no-u-turn, and no-parking, as shown in \figurename~\ref{fig:traffic_signs}.
    
    % Figure --------------------------------------
    \begin{figure*}[!t]
      \centering
      \includegraphics[width=\textwidth]{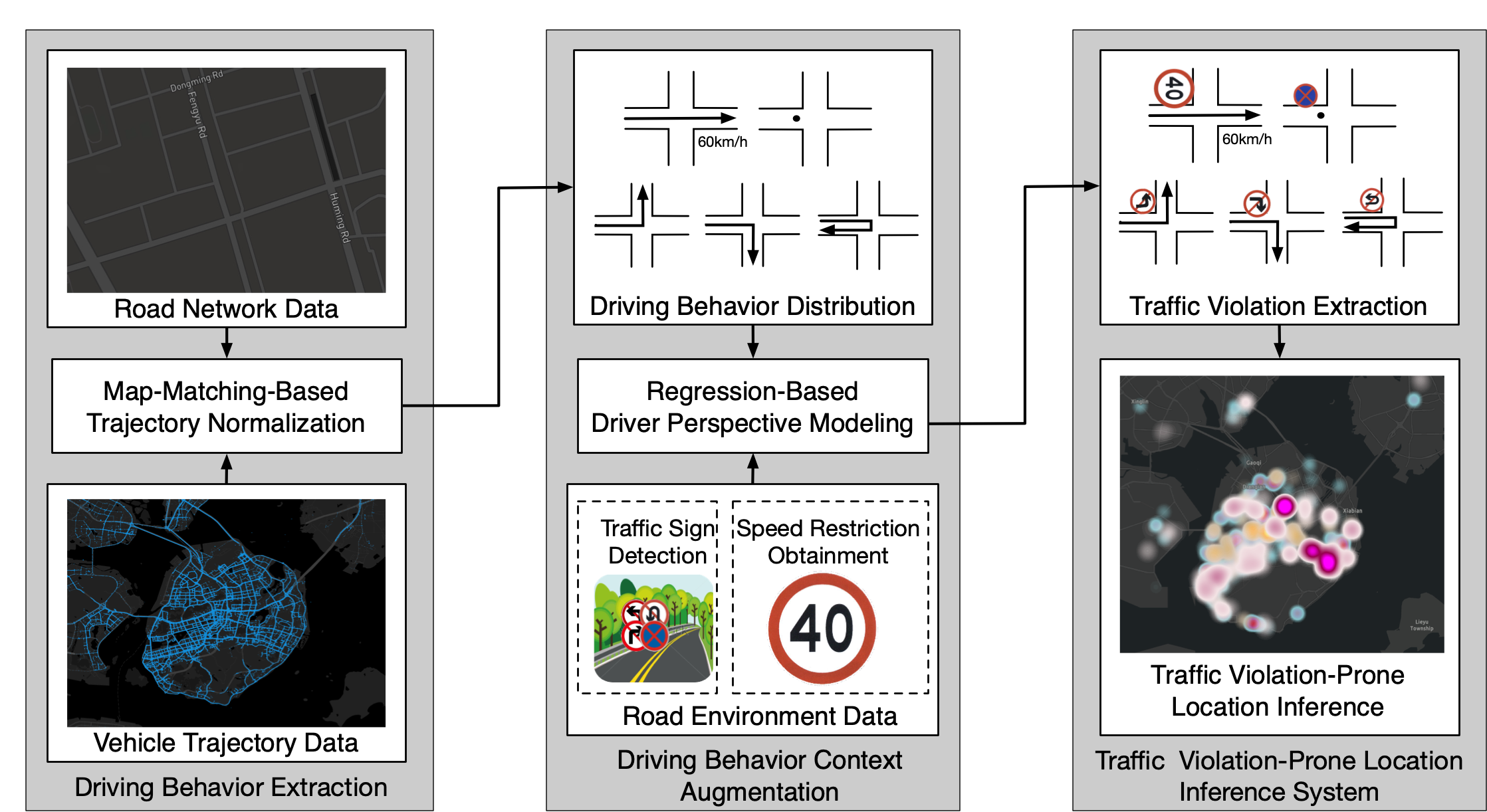}
      \caption{Framework Overview.}
      \label{fig:framework}
      \vspace{-0.5cm}
    \end{figure*}
    % \vspace{-0.8cm}
   
  \subsection{System Framework Overview}

    As shown in \figurename~\ref{fig:framework}, the framework consists of three phases, i.e., driving behavior extraction, driving behavior context augmentation and traffic violation-prone location inference system. We briefly elaborate on the whole process as follows.

    In the driving behavior extraction phase, we normalize the trajectory data based on the road network using map matching methods. Then we extract parking behaviors from the normalized data by static points and extract turning behaviors from the road segment trajectory. In the driving behavior context augmentation phase, we model driver perspectives based on regression and retrieve corresponding street view pictures and other traffic restriction information. Then we train a traffic sign detection model to detect traffic signs in street view pictures and obtain speed restrictions from real-time navigation service providers. In the traffic violation-prone location inference system phase, we first match driving behaviors with those restrictions to get the traffic violation distribution. Then we extract the spatiotemporal patterns of traffic violations to infer the traffic violation-prone locations and build a traffic violation-prone location inference system.

% ==========================================================
\section{Driving Behavior Extraction}

  In this section, our goal is to extract average velocities and driving behaviors, i.e., parking, left turn, right turn, and u-turn from crowd-sensed and large-scale vehicle GPS trajectories. However, the GPS readings are usually noisy and nonuniform due to many factors such as poor conditions of GPS devices and different sampling rates \cite{harati2007automatic}, and it is non-trivial to extract driving behaviors directly from the trajectories without the information about road network. To address these challenges, we first employ a map matching method to normalize trajectory data into road segment trajectories, and then extract the driving behaviors, i.e., parking, left turn, right turn, and u-turn from the normalized trajectories. 

  % There are several challenges in practice: (1) GPS readings are noisy. In general, the Civilian GPS devices have an error of about 10 meters in reporting positions \cite{harati2007automatic}, and there are several factors that will influence the precision of GPS records. For example, the poor conditions of receiving devices and sending devices may cause noisy trajectories. (2) Even with clean GPS trajectory data, it is not trivial to extract driving behaviors directly from the trajectories. As aforementioned, due to the road network diversity and different sampling frequencies, we cannot extract correct driving behaviors directly from GPS trajectories.

  % To address the above challenges, we propose a two-step process to extract the driving behaviors from GPS trajectories. First, we employ map-matching methods to normalize trajectory data into road segment trajectories. Second, we extract the average velocity and driving behaviors, i.e., parking, left turn, right turn, and u-turn from the normalized trajectories.

  % ---------------------------------------------
  \subsection{Map-Matching-Based Trajectory Normalization}
  %implemented by Open Source Routing Machine(OSRM) routing engine\cite{luxen2011real}, 

    % Figure --------------------------------------
    \begin{figure*}[!t]
      \centering
      \subfloat[]{
            \begin{minipage}[c]{.22\linewidth}
                \centering
                \includegraphics[width=\textwidth]{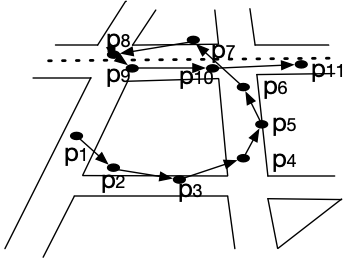}
            \end{minipage}
        \label{fig:map_matching_1}
        }
        \quad
        \subfloat[]{    
            \begin{minipage}[c]{.22\linewidth}
                \centering
                \includegraphics[width=\textwidth]{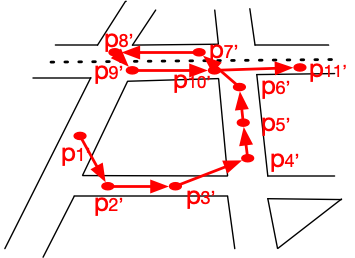}
            \end{minipage}
            \label{fig:map_matching_2}
        }
      \quad
        \subfloat[]{ 
            \begin{minipage}[c]{.22\linewidth}
                \centering
                \includegraphics[width=\textwidth]{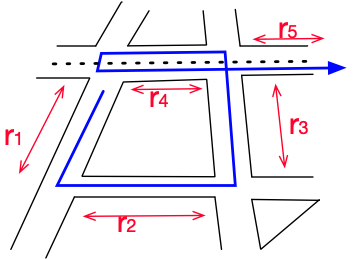}
            \end{minipage}
            \label{fig:map_matching_3}
        }
        \quad
        \subfloat[]{ 
            \begin{minipage}[c]{.22\linewidth}
                \centering
                \includegraphics[width=\textwidth]{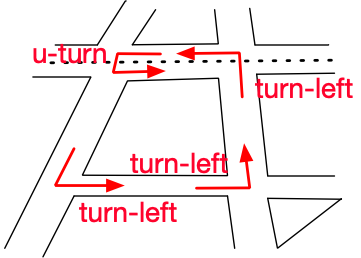}
            \end{minipage}
            \label{fig:map_matching_4}
        }
        \caption{An illustration of map-matching-based trajectory normalization. (a) A GPS point trajectory. (b) A normalized GPS point trajectory. (c) A normalized road segment trajectory. (d) A turning behavior trajectory.}
        \label{fig:map_matching}
        \vspace{-0.5cm}
    \end{figure*}

  The purely geometric map matching methods are sensitive to measurement noise and sampling rate, while the trajectory data are usually noisy and nonuniform in real-world deployment. Therefore, in this paper, a HMM-based map matching algorithm which addresses this issue is used to find the most likely road route represented by a timestamped sequence of latitude/longitude pairs \cite{newson2009hidden}.

  We first preprocess the raw vehicle trajectories by removing duplicate and abnormal points and reconstructing trajectories. In this algorithm, the states of the HMM are the individual road segments, denoted as $r_i$, $i=1,...,m$, and the state measurements are the noisy vehicle location measurements. The goal is to match each latitude/longitude location measurement $z_t$ with the proper road segment. 

  The emission probability for each road segment $r_i$ and each location measurement $z_t$ is $p(z_t|r_i)$, which gives the likelihood that $z_t$ would be observed if the vehicle were actually on road segment $r_i$, and the closest point on the road segment is denoted as $x_{t,i}$, and direct distance (the great circle direct distance on the surface of the earth between two GPS points) between the measured point and the candidate match is $\Vert z_t-x_{t,i}\Vert_{d}$. The intuition is that road segments farther from the measurement are less likely to have produced the measurement. The GPS noise are modeled as zero-mean Gaussian \cite{van2007gnss}, i.e.,
  \begin{equation}\label{equ:gaussion}
  p(z_t|r_i)=\frac{1}{\sqrt{2\pi}\sigma_z}e^{-0.5(\frac{\Vert z_t-x_{t,i}\Vert_d}{\sigma_z})^2}
  \end{equation}
  where $\sigma_z$ is the standard deviation of GPS measurements, and the points within $2\sigma_z$ of the previously included point are removed. $\sigma_z$ is estimated by the median absolute deviation \cite{newson2009hidden}, i.e., 
  \begin{equation}\label{equ:sigma}
  \sigma_z=1.4826\times\text{median}_t(\Vert z_t-x_{t,i}\Vert_d)
  \end{equation}
  
  The initial state probabilities is started at the first measurement, $\pi_i=p(z_1|r_i)$, $i=1,...,m$. Also, we do not consider matching to road segments that are quite distant from the measurement. The measurement probability from a road segment that is more than 200 meters away from $z_t$ is set to zero, which helps reduce the number of candidate matches.

  The transition probabilities give the probability of a vehicle moving between the candidate road matches at $t$ and $t+1$. Specifically, for the next measurement $z_{t+1}$ and candidate road segment $r_j$, the corresponding point is $x_{t+1,j}$. We compute the direct distance between $z_t$ and $z_{t+1}$, $\Vert z_t-z_{t+1}\Vert_d$ and the route distance (the shortest length of road segments between two GPS points on the road) between $x_t$ and $x_t+1$, $\Vert x_{t,i}-x_{t+1,j}\Vert_r$. With the intuition that these two distances will be about the same for correct matches, and the absolute values of these two distances from the correct matches fits well to an exponential probability distribution given by \cite{newson2009hidden}
  \begin{equation}\label{equ:transition_probabilities}
  p(d_t)=\frac{1}{\beta}e^{-\frac{d_t}{\beta}}
  \end{equation}
  where $d_t=\vert\Vert x_{t,i}-x_{t+1,j}\Vert_r-\Vert z_t-z_{t+1}\Vert\vert_d$, and $\beta$ is estimated by a robust estimator suggested by Gather and Schultze \cite{gather1999robust}, i.e.,
  \begin{equation}\label{equ:beta}
  \beta=\frac{1}{ln(2)}\text{median}_t(\vert(\Vert z_t-z_{t+1}\Vert_d-\Vert x_{t,i}-x_{t+1,j}\Vert_r)\vert)
  \end{equation}
  We set $p(d_t)$ to zero when $d_t$ is greater than 2000 meters and terminate the search for a route. And if a calculated route would require the vehicle to exceed a speed of 180 kilometers per hour, we set its probability to zero. 
  With the measurement probabilities and transition probabilities, the Viterbi algorithm \cite{forney1973viterbi} is used to compute the best path.

    % % Figure --------------------------------------
    % \begin{figure}
    %     \centering
    %     \includegraphics[width=.4\textwidth]{robustness.png}
    %     \caption{\hl{The performance of the map-matching method under different sampling periods and noise on the location measurements \cite{newson2009hidden}.}}
    %     \label{fig:robustness}
    % \end{figure}

  % \hl{\figurename{~\ref{fig:robustness}} provided by \cite{newson2009hidden} shows the performance of this map-matching algorithm under different sampling periods and noise on the location measurements. The algorithm shows robustness to sampling periods less than 90 seconds, and if the sampling period is in reasonable range, the algorithm can achieve good performance even with the noise standard deviation up to 40 meters.}

  After map matching, the GPS point trajectory $traj_p=(id,P_s)$ is normalized into road segment trajectory $traj_g=(id,R_s)$. For example, as shown in \figurename~\ref{fig:map_matching_1}, there is a GPS point trajectory $traj_{p}=(i,P_{s})$, where $P_{s}=\{p_1\to p_2\to p_3\to p_4\to p_5\to p_6\to p_7\to p_8\to p_9\}$. After map matching, we get the corresponding locations of each GPS points on the road segment, as shown in \figurename~\ref{fig:map_matching_2}. The mapped point $p_1'$ is on road segment $r_1$, $p_2'$ and $p_3'$ are on road segment $r_2$, $p_4'$ to $p_6'$ are on road segment $r_3$, $p_7'$ to $p_{10}'$ are on $r_4$, $p_{11}'$ is on $r_5$. Thus we can get the corresponding road segment trajectory $traj_{g}=(i,R_{s})$, where $R_{s}=\{r_1\to r_2\to r_3\to r_4\to r_4\to r_5\}$, as shown in \figurename~\ref{fig:map_matching_3}.    

  \subsection{Driving Behavior Extraction}
  Based on the normalized trajectories, we further extract vehicles' average velocities on roads, the turning behaviors in road intersections, and the parking behaviors on road segments from the normalized trajectories.

    \subsubsection{Turning Behavior Extraction} From the normalized road segment trajectories, we extract the turning behaviors. $bb$ and $ba$ are the directions of the vehicle before and after the turning behavior, which are the clockwise angles from the North. If $ba-bb=0^{\circ}$, it is a straight behavior, which is not considered in this paper; if $0^{\circ}<ba-bb<160^{\circ}$, it is a right-turn behavior; if $160^{\circ}\leq ba-bb\leq 200^{\circ}$, it is a u-turn behavior; if $200^{\circ}<ba-bb<360^{\circ}$, it is a left-turn behavior. As shown in \figurename~\ref{fig:map_matching}, from $r_1$ to $r_2$, there is a left-turn behavior; from $r_2$ to $r_3$, there is a left-turn behavior; from $r_3$ to $r_4$, there is a left-turn behavior; from $r_4$ to $r_4$, there is a u-turn behavior; from $r4$ to $r_5$, it is a straight behavior. Therefore we can get a turning behavior trajectory, $\{\text{left-turn}\to \text{left-turn}\to \text{left-turn}\to \text{u-turn}\}$, as shown in \figurename~\ref{fig:map_matching} (d). 

    \subsubsection{Parking Behaviors Extraction}

    % Figure --------------------------------------
    \begin{figure}[!t]
      \centering
      \includegraphics[width=.4\textwidth]{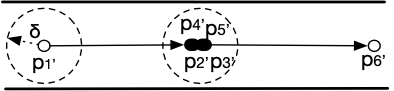}
      \caption{An illustration of extracting the parking behavior.}
      \label{fig:static_points}
    \end{figure}

    We extract the parking behaviors in the normalized GPS point trajectories with a sliding-window-based method \cite{chen2016container}. More specifically, for a normalized GPS point trajectory $traj_p=(i,P_s)$, where $P_s=\{p_1'\to p_2'\to ...\to p_n'\}$, we extract every parking sequence $p_m'\to p_{m+1}'\to ...\to p_{m+k}'(1\leq m<n, 1\leq k\leq n-m)$ in which the average speed between the first point and any other points is less than a small threshold $\delta$ \cite{chen2016container}, i.e.,
    \begin{equation}\label{equ:parking_behaviors}
    \forall m\leq i<m+k, \frac{dist(p_i',p_{i+1}')}{\bigtriangleup t}<\delta
    \end{equation}
    where $dist(p_i',p_{i+1}')$ is the route distance between $p_i'$ and $p_{i+1}'$ and $\bigtriangleup t=|p_i'.t-p_{i+1}'.t|$ is the time difference of $p_i'$ and $p_{i+1}'$. We use a sliding-window with adaptive size along the trajectory to find such parking behaviors. In particular, we dynamically extend the window size by adding new points until the newly-formed sequence violates requirement \ref{equ:parking_behaviors}. For example, as shown in \figurename~\ref{fig:static_points}, for the trajectory $p_1'\to p_2'\to p_3'\to p_4'\to p_5'\to p_6'$, we start by creating a window consisting of the first two points ($p_1',p_2'$ in this case), and check whether the average speed between $p_1'$ and $p_2'$ is less than $\delta$. Since $\frac{dist(p_1',p_2')}{|p_1'.t-p_2'.t|}>\delta$, we discard this window, and slide the window to start over from the end point($p_2'$), and create a new window ($p_2',p_3'$). We keep this window for $\frac{dist(p_2',p_3')}{|p_2'.t-p_3'.t|}<\delta$ and repeat the procedure for the next adjacent points until the speed constraint is violated. Finally, we obtain a sequence containing a set of consecutive points $p_2'\to p_3'\to p_4'\to p_5'$. 

    We map each parking sequence $p_m'\to p_{m+1}'\to...\to p_{m+k}'$ extracted from the GPS trajectories to a parking behavior $pk=(id,lat,lng,st,et)$. The $id$ of $pk$ is the $id$ of the trajectory where the parking behavior is extracted, $lat$ is the average latitude of the points in the parking sequence, $lat=\frac{\sum_{i=m}^{m+k}p_i'.latitude}{k+1}$, $lng$ is the average longitude of the points in the parking sequence, $lng=\frac{\sum_{i=m}^{m+k}p_i'.longitude}{k+1}$, $st$ is start time of the parking behavior, $st=p_m'.t$, $et$ is the end time of the parking behavior, $et=p_{m+k}'.t$.

    \subsubsection{Average Velocity Extraction} After trajectory normalization, we can get sequences of road intersections, from which we extract the subsequences. The intersections from the same subsequence are of the same road. For each subsequence, we calculate the average speed through dividing the route distance between the head intersection and the rear intersection by the difference of the time. More specifically, for a road intersection sequence $p_1\to p_2\to p_3\to ...\to p_n$, where $n$ is the length of the sequence. If $p_i$, $p_{i+1}$ and $p_{i+2}$ are of the same road, we extract the subsequences $p_i\to p_{i+1}\to p_{i+2}$, and the average speed $v=\frac{dist(p_{i},p_{i+2})}{\bigtriangleup t}$, where $dist(p_{i},p_{i+2})$ and $\bigtriangleup t$ are the route distance and duration between $p_{i}$ and $p_{i+2}$.

% ==========================================================
\section{Driving Behavior Context Augmentation}

  After extracting driving behaviors, we need to know the contexts of driving behaviors to identify whether a driving behavior is illegal or not. To this end, we model the driver perspectives based on regression and get the corresponding street view pictures to detect the no-left-turn, no-right-turn, no-u-turn, and no-parking signs. 

  \subsection{Driver Perspective Modeling}

    Since vehicle trajectory data we use in this paper are extremely large-scale, the GPS points in the trajectories can almost cover the urban road network completely. Therefore, after getting the driving behavior database $DB=(PK,TN)$, we can detect almost all the road intersections in the city by extracting the distinct location of turning behaviors. The no-left-turn, no-right-turn, and no-u-turn traffic signs are erected near the road intersections to guide drivers. And the no-parking signs are often erected on the road intersections to inform drivers whether they can park on the following road segments.

\begin{figure}
        \begin{minipage}[c]{.41\linewidth}
        \centering
        \includegraphics[width=\textwidth]{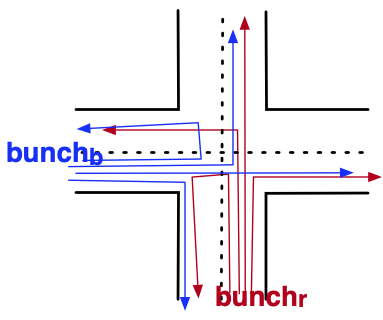}
        \caption{An example of two $bunches$ of trajectories.}
        \label{fig:bunch}
        \end{minipage}
        \quad
        \begin{minipage}[c]{.43\linewidth}
        \centering
        \includegraphics[width=\textwidth]{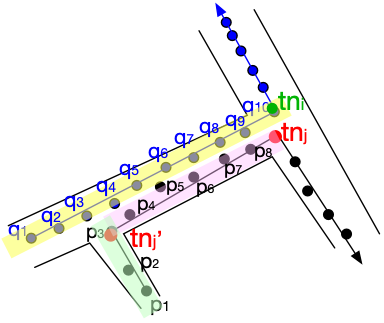}
        \caption{An illustration of the driver perspective modeling.}
        \label{fig:timespan}
        \end{minipage}
        \vspace{-0.8cm}
    \end{figure}

    For each intersection, there are several different kinds of turning behaviors from different directions. We define those turning behaviors at the same road intersection from the same direction as a $bunch=(lat,lng,bb)$, where $lat,lng$ are the latitude and the longitude of the intersection, $bb$ is the bearing before the behaviors. Thus for the turning behaviors from one $bunch$, their spatiotemporal contexts before the behaviors should be the same. As shown in \figurename~\ref{fig:bunch}, the behaviors drew in red are a $bunch$, denoted as $bunch_r$ and those in blue are another $bunch$, denoted as $bunch_b$. 

    For each $bunch_i=(lat_i,lng_i,bb_i)$, we select the corresponding GPS point trajectories in $bunch_i$, and for each trajectory $j$, the corresponding turning behavior at this road intersection is $tn_j$, and we select a segment of the trajectory $traj_{pj}=\{p_k\to p_{k+1}\to...\to p_{k+m}\}$ before $tn_j$, the time span of which is limited by a threshold $\theta$. Moreover, in order to avoid the GPS points on the branch road, the start time of the trajectory segment should be no earlier than the former turning behavior $tn_j'$ of $tn_j$, i.e. ,
    \begin{equation}\label{equ:time_threshold}
    tn_j.t-p_k.t<\theta, p_{k+m}\leq tn_j.t, tn_j'.t\leq p_k.t\leq tn_j.t
    \end{equation}

    % Figure --------------------------------------

    For example, as shown in \figurename~\ref{fig:timespan}, suppose that there are two GPS trajectories in the $bunch$ at this road intersection, according to \ref{equ:time_threshold}, we select two trajectory segments, $traj_{p1}=\{q_1\to q_2\to q_3\to q_4\to q_5\to q_6\to q_7\to q_8\to q_9\to q_{10}\}$ and $traj_{p2}=\{p_1\to p_2\to p_3\to p_4\to p_5\to p_6\to p_7\to p_8\}$, where $tn_i.t-q_1.t<\theta$, $q_{10}\leq tn_i.t$, $tn_j.t-p_1.t<\theta$ and $p_8\leq tn_j.t$. However, for $traj_{p2}$, there is a right-turn behavior $tn_j'$ before $tn_j$, $p_1.t<tn_j'.t$ and $p_2.t<tn_j'.t$, so we drop points $p_1$ and $p_2$. Finally, the trajectory segments selected are $q_1\to q_2\to q_3\to q_4\to q_5\to q_6\to q_7\to q_8\to q_9\to q_{10}$ and $p_3\to p_4\to p_5\to p_6\to p_7\to p_8$, respectively.

    % Figure --------------------------------------
    \begin{figure}
      \centering
      \subfloat[]{
            \begin{minipage}[c]{.3\linewidth}
                \centering
                \includegraphics[width=.8\textwidth]{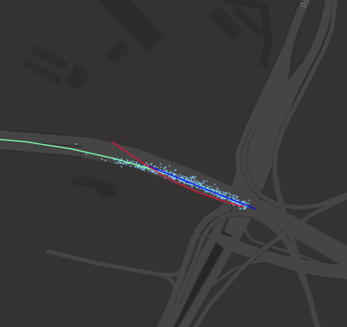}
            \end{minipage}
        \label{fig:regression_1}
        }
        \subfloat[]{    
            \begin{minipage}[c]{.3\linewidth}
                \centering
                \includegraphics[width=.8\textwidth]{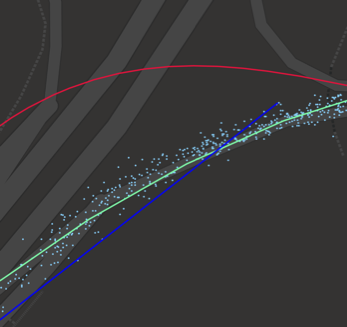}
            \end{minipage}
            \label{fig:regression_2}
        }
        \subfloat[]{ 
            \begin{minipage}[c]{.3\linewidth}
                \centering
                \includegraphics[width=.8\textwidth]{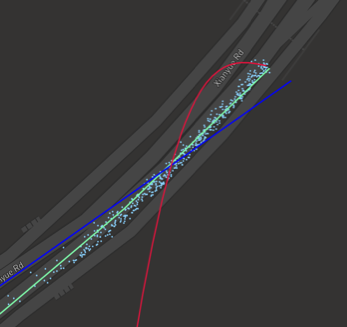}
            \end{minipage}
            \label{fig:regression_3}
        }
        \caption{The results of linear regression (the blue lines), spline regression (the red lines) and cubic polynomial regression (the green lines).}
        \label{fig:regression}
    \end{figure}

    Since the trajectory data are extremely massive, we select a large number of trajectory segments that can describe contexts before turning behaviors at the road intersections. For each turning behaviors $tn_i$, we denote the GPS points in the corresponding trajectory segments as a set $P_i=\{(x_1,y_1),(x_2,y_2),...,(x_m,y_m)\}$, where $m$ is the number of the points, we model the shape of the lane segment by regression. We select five typical road segments and try three kinds of regression, linear regression, spline regression, and cubic polynomial regression. The result shows that cubic polynomial regression achieves the best performance, as shown in \figurename{~\ref{fig:regression}}. 

    Therefore for $(x_i,y_i)(i=1,...,m)$, we can get a curve $h(x)=\theta_0+\theta_1x+\theta_2x^2+\theta_3x^3$ satisfied that \cite{williams2016quantifying, lawson1995solving}, 
    \begin{equation}
    \theta_0,\theta_1,\theta_2,\theta_3=\mathop{\arg\min}_{\theta_0,\theta_1,\theta_2,\theta_3}\frac{1}{2m}\sum_{i=1}^m(h(x_i)-y_i)^2
    \end{equation}
    The regression curve segment can be regarded as a virtual driver's route before the driving behavior, representing all drivers of the selected trajectories. 

  \subsection{Driving Behavior Context Augmentation}

\begin{figure}[!t]
      \vspace{-0.3cm}
      \centering
        \subfloat[]{ 
            \begin{minipage}[c]{.3\linewidth}
                \centering
                \includegraphics[width=\textwidth]{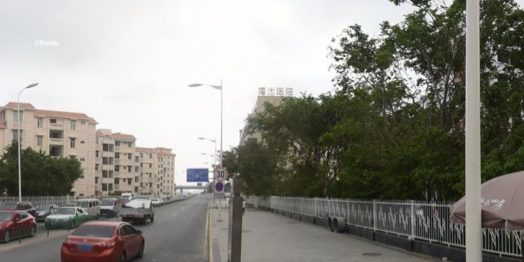}
            \end{minipage}
            \label{fig:street_view_sequence_3}
        }
        \subfloat[]{
            \begin{minipage}[c]{.3\linewidth}
                \centering
                \includegraphics[width=\textwidth]{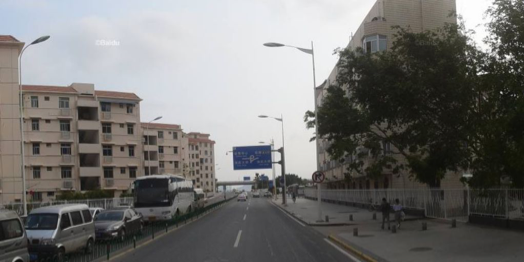}
            \end{minipage}
        \label{fig:street_view_sequence_4}
        }
        \subfloat[]{    
            \begin{minipage}[c]{.3\linewidth}
                \centering
                \includegraphics[width=\textwidth]{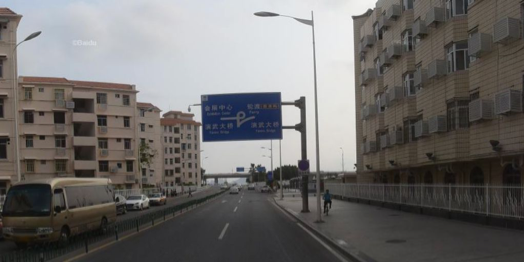}
            \end{minipage}
            \label{fig:street_view_sequence_5}
        }
        \caption{An example of a sequence of street view pictures.}
        \label{fig:svseq}
        \vspace{-0.5cm}
    \end{figure}

    We pick several points as a point sequence on the regression curve segment uniformly and normalized them by map matching, regard the tangent of the curve at those points as the direction of the vehicle at these locations. Therefore we can get the corresponding street view picture sequence, through which we can restore what drivers have seen before those road intersections. For example, \figurename~\ref{fig:svseq} shows a sequence of street view pictures. In addition, we also collect the speed restrictions of each road from real-time navigation service providers. 

% ==========================================================
\section{Traffic Violation-Prone Location Inference System}
  
  In this section, our goal is to infer the traffic violation-prone locations in cities. However, since the intelligent information management system have not been implemented in many developing or developed cities, the traffic management departments usually do not have a comprehensive knowledge about the traffic restriction distribution in the whole city. Therefore, we obtain the traffic rule information through detecting traffic signs in street view pictures so as to identify traffic violations. Then we extract the spatiotemporal patterns of traffic violations to infer traffic violation-prone locations. 

  \subsection{Traffic Sign Detection}

    From driver perspectives, we can find no-left-turn signs, no-right-turn signs, no-u-turn signs, and no-parking signs erected at the side of or above the roads to give instructions to drivers. After modeling driver perspectives, we then detect these four categories of traffic signs in the spatiotemporal context around the road intersections to help infer the illegal turning behaviors.

    % Figure --------------------------------------
    \begin{figure}
    % \vspace{-0.3cm}
      \centering
      \includegraphics[width=.49\textwidth]{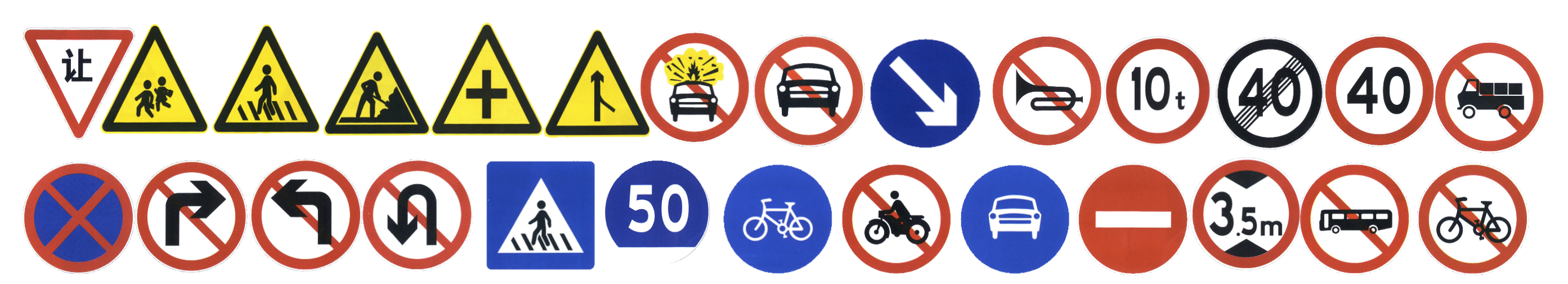}
      \caption{The types of traffic signs selected from the Chinese traffic-sign benchmark.}
      \label{fig:signs}
      \vspace{0cm}
    \end{figure}

    % Figure --------------------------------------
    \begin{figure}[!t]
    \vspace{-0.3cm}
      \centering
      \subfloat[]{
            \begin{minipage}[c]{.47\linewidth}
                \centering
                \includegraphics[width=\textwidth]{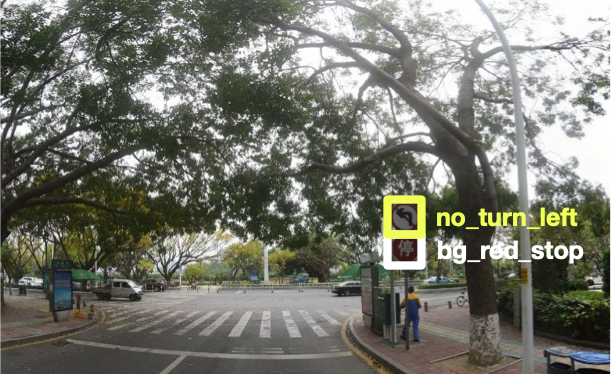}
            \end{minipage}
        \label{fig:training_pictures_1}
        }
        \subfloat[]{    
            \begin{minipage}[c]{.47\linewidth}
                \centering
                \includegraphics[width=\textwidth]{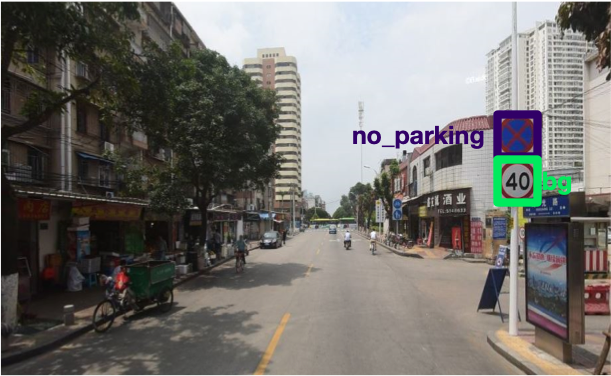}
            \end{minipage}
            \label{fig:training_pictures_2}
        }\\
        \vspace{-0.3cm}
        \subfloat[]{    
            \begin{minipage}[c]{.47\linewidth}
                \centering
                \includegraphics[width=\textwidth]{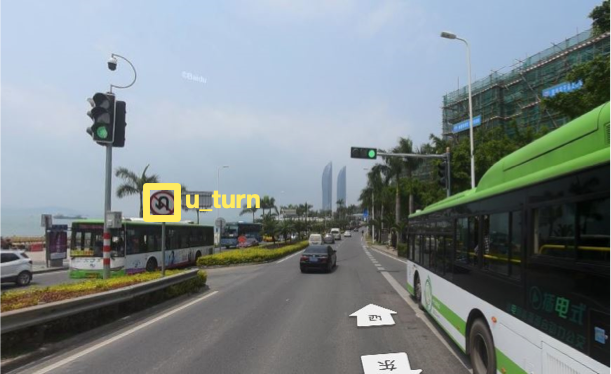}
            \end{minipage}
            \label{fig:training_pictures_3}
        }
        \subfloat[]{    
            \begin{minipage}[c]{.47\linewidth}
                \centering
                \includegraphics[width=\textwidth]{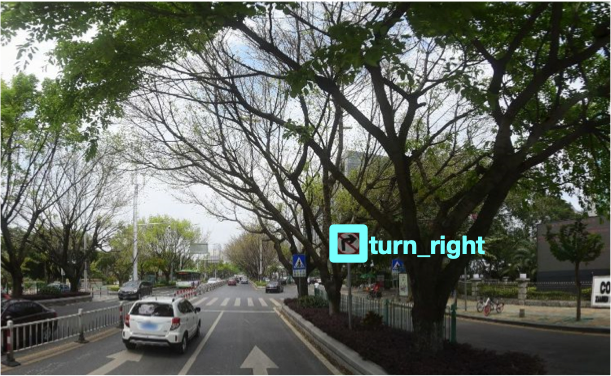}
            \end{minipage}
            \label{fig:training_pictures_4}
        }
        \caption{Examples of training pictures. (a) A no-left-turn sign and a negative sample. (b) A no-parking sign and a negative sample. (c) A no-u-turn sign. (d) A no-right-turn sign.}
      \label{fig:training_pictures}
      % \vspace{-0.5cm}
    \end{figure}

    The state-of-art object detection system YOLOv3 \cite{redmon2018yolov3} has achieved great performance both in accuracy and time. It has a few incremental improvements on YOLOv2 \cite{redmon2017yolo9000}, such as using independent logistic classifiers instead of softmax, adding shortcut connections, and concatenating feature maps with upsampled features, which helps it make great progress in small object detection and has strong generalization ability. Thus we choose YOLOv3 as the backbone network to train the traffic sign detection model with transfer learning \cite{pan2009survey}.

    We first use the Chinese traffic-sign benchmark created by Zhu et al. \cite{Zhe_2016_CVPR} to pre-train a traffic sign detection model. \figurename~\ref{fig:signs} shows the types of traffic signs selected from the benchmark. Then we fine-tune the model based on a small dataset consisting of four categories of traffic signs, i.e., no-turn-left, no-turn-right, no-u-turn, and no-parking, collected from the targeted city. The small dataset is consistent with the street view picture dataset, as shown in \figurename~\ref{fig:training_pictures}. Therefore, if we plan to transfer the model to another city, we only need to collect a small dataset consisting of the interested categories of traffic signs from the new city and fine-tune the pre-trained model rather than collect a large dataset and re-train a new detection model.

  \vspace{-0.3cm}
  \subsection{Traffic Violation Identification}

    \subsubsection{Illegal Turning} If there exists the corresponding traffic sign in the street view picture sequence before the turning behavior, such as a no-left-turn sign for a left-turn behavior, this behavior can be identified as an illegal turning behavior. We denote the no-turning road intersections in the city as $L=\{l_1,l_2,...,l_m\}$, where $l_i(i=1,...,m)$ is the distribution of illegal turning behaviors on each unique intersection.

    \subsubsection{Illegal Parking} From the traffic sign database constructed above, we can get the locations of the no-parking traffic signs. After mapping these locations to the nearest road segments, we get the list of no-parking road segments. If the corresponding road segment of a parking behavior is in the no-parking road segment list and the direct distance between the no-parking sign and the parking behavior is less than a threshold $\zeta$, the parking behavior can be identified as an illegal parking behavior. We denote the no-parking segments in the city as a set $B=\{b_1,b_2,...,b_n\}$, where $b_i(i=1,...,n)$ is the distribution of illegal parking behaviors on each unique road segment.

    \subsubsection{Speeding} As for speeding, we first collect road speed restrictions from Gaode Map Open Platform \footnote{https://lbs.amap.com} and get the list of speed restrictions on the roads. Then we compare the average speed with the speed restrictions of the corresponding roads. If the average speed exceeds the speed restriction on the road, it can be identified as a speeding behavior. We denote the roads with speed restrictions in the city as $S=\{s_1,s_2,...,s_l\}$, where $l$ is the number of roads, $s_i(i=1,...,l)$ is the distribution of speeding behaviors on each unique road.

  \subsection{Traffic Violation-Prone Location Inference}
    After identifying traffic violations, we extract the spatiotemporal patterns of traffic violations to infer the traffic violation-prone locations.

    For each traffic violation-prone location candidate, we aggregate the traffic violations related to this location. More specifically, We aggregate different illegal turning behaviors on the same intersection, parking behaviors on the same road segment and speeding behaviors on the same road.

    For each traffic violation-prone location candidate $r_n$ ($r_n\in L\cup B\cup S$), we aggregate its hourly traffic violations to build the temporal profile, i.e.,
    \begin{align}
      \Phi(r_n) = [tv_1,tv_2...,tv_{H}]
    \end{align}
    where $ tv_i (i = 1,2...,H) $ is the number of traffic violations of the $ i^{th} $ hour and $H$ is the number of hours. Then for each traffic violation-prone location candidate, we aggregate and average hourly traffic violations in the dataset over a typical day to determine the threshold of each hour in a day to infer traffic violation-prone locations, i.e.,
    \begin{align}
      \overline{\Phi(r_n)} = [\overline{tv_1},\overline{tv_2}...,\overline{tv_{24}}]
    \end{align}
    where $ \overline{tv_i} (i = 1,2...,24) $ is the average number of aggregated traffic violations of the $ i^{th} $ hour in a day.

    For each hour $j (j=1,2,...,24)$, we set the threshold of this hour as the average number of the traffic violations plus the double standard deviation of the traffic violations in the city \cite{pukelsheim1994three, wheeler1992understanding, czitrom1997statistical}, i.e.,
    \begin{align}
    \mu_j=\frac{\sum_{i=1}^N \overline{tv_j}^i}{N}, \sigma_j=\sqrt{\frac{\sum_{i=1}^N(\overline{tv_j}^i-\mu_j)^2}{N}},th_j=\mu_j+2\sigma_j
    \end{align}
    % \begin{align}
    
    % \end{align}
    where $N$ is the number of traffic violation-prone location candidates, $\overline{tv_j}^i$ is the typical traffic violation of $r_i$ at hour $j$. Therefore, we can get the traffic violation-prone locations for each hour.

    % Figure --------------------------------------
    % \begin{figure}[!t]
    %   \centering
    %   \includegraphics[width=.4\textwidth]{traffic_violation_dis.png}
    %   \caption{The number of traffic violations in Chengdu in November, 2016}
    %   \label{fig:vio_dis}
    % \end{figure}

% ==========================================================
\section{Evaluation}

  In this section, we evaluate the performance of the proposed method based on two crowd-sensed, large-scale, and real-word trajectory datasets from two cities in China, Xiamen and Chengdu, respectively \footnote{Sample datasets and codes can be found at: https://github.com/zhihanjiang/iTV}. We first introduce the datasets and experiment settings. Then we present the evaluation results and a traffic violation-prone location inference system. Finally, we conduct several case studies. 

    % % Figure --------------------------------------
    % \begin{figure}[!t]
    %   \centering
    %   \subfloat[]{
    %             \centering
    %             \includegraphics[width=.23\textwidth]{trajectory_1.png}
    %     \label{fig:trajectory_1}
    %     }
    %     \subfloat[]{    
    %             \centering
    %             \includegraphics[width=.23\textwidth]{trajectory_2.png}
    %         \label{fig:trajectory_2}
    %     }
    %     \caption{The GPS trajectories in Xiamen and Chengdu. (a) GPS trajectories in Xiamen. (b) GPS trajectories in Chengdu.}
    %   \label{fig:trajectories}
    % \end{figure}

  \subsection{Dataset Description} 

      \begin{table}[!t]
      \renewcommand{\arraystretch}{1.3}
      \caption{Vehicle Trajectory Dataset Description}
      \label{table:traj}
      \centering{}
      \begin{tabularx}{.45\textwidth}{ @{} m{2.7cm} m{5cm}  @{} } 
        \toprule
        \bfseries City          &\bfseries Xiamen    \\
        \midrule
        Duration & 09/01/2016 00:00:00 - 09/30/2016 23:59:59  \\
        Record &  352,300,768 (5378 vehicles) \\
        Latitude (WGS84$^a$)    &   $24.369406^{\circ}$N - $24.619351^{\circ}$N  \\
        Longitude (WGS84)   &  $117.990364^{\circ}$E - $118.265022^{\circ}$E\\
        \midrule
        \bfseries City          &\bfseries Chengdu     \\
        \midrule
        Duration &  11/01/2016 00:00:00 - 11/30/2016 23:59:59 \\
        Record &   1,096,608,443 (6,096,022 orders)\\
        Latitude (WGS84$^a$)    & $30.655297^{\circ}$N - $30.730203^{\circ}$N \\
        Longitude (WGS84)   & $104.039652^{\circ}$E - $104.127062^{\circ}$E\\
        \bottomrule
      \end{tabularx}\\
      \footnotesize{$^a$ World Geodetic System 1984, the reference system for the Global Positioning System (GPS)}
      \end{table}

    \subsubsection{\textbf{Taxi Trajectory Data in Xiamen and Car-Hailing Trajectory Data in Chengdu}} The taxi trajectory data in Xiamen are provided by Xiamen Traffic Management Department. After a data cleansing process that removes invalid records. We obtain the taxi trajectories of 5,486 vehicles in Xiamen, Fujian Province, in China. The car-hailing trajectory data are provided by GAIA Open Dataset\footnote{https://outreach.didichuxing.com/research/opendata/} from DiDiChuxing, the largest online car-hailing service provider in China, which handles around 11 million orders per day all over China\footnote{https://www.didiglobal.com}. After a data cleansing process that removes invalid records. We obtain the vehicle trajectories of 6,096,022 orders. These two datasets are both crowd-sensed and large-scale. The detailed summary of these two datasets are shown in TABLE. \ref{table:traj}.

    \subsubsection{\textbf{Speed Restriction Information}}
    We collect the speed restrictions on 150 roads in Xiamen and 281 roads in Chengdu from Gaode Map Open Platform, as shown in TALE. \ref{table:sp_limits}.

    % % Figure --------------------------------------
    % \begin{figure}
    %   \centering
    %   \subfloat[Speed restriction distribution on roads in Xiamen.]{
    %             \centering
    %             \includegraphics[width=.25\textwidth]{sp_limits_1.png}
    %     \label{fig:sp_limits_1}
    %     }
    %     \subfloat[Speed restriction distribution on roads in Chengdu.]{    
    %             \centering
    %             \includegraphics[width=.25\textwidth]{sp_limits_2.png}
    %         \label{fig:sp_limits_2}
    %     }
    %     \caption{The speed restriction distributions on roads in Xiamen and Chengdu.}
    %   \label{fig:sp_limits}
    % \end{figure}

      \begin{table}[!t]
      \renewcommand{\arraystretch}{1.3}
      \caption{The number of different speed restrictions on roads in Xiamen and Chengdu.}
      \label{table:sp_limits}
      \centering{}
      \begin{tabularx}{.45\textwidth}{ @{} m{2.5cm} m{0.3cm} m{0.3cm} m{0.3cm} m{0.3cm} m{0.3cm} m{0.3cm} m{0.3cm} @{} } 
        \toprule
        \bfseries Speed Limit (km/h)  &\bfseries 5 &\bfseries 30 &\bfseries 40 &\bfseries 50&\bfseries 60&\bfseries 70 &\bfseries 80     \\
        \midrule
        Xiamen & 5 &2 & 89 &25 &25&3&1 \\
        Chengdu & 6 &17 &220 &8 &28 &1 &1\\
        \bottomrule
      \end{tabularx}
      \end{table}

    \subsubsection{\textbf{Street View Pictures for Traffic Sign Detection Model Training}} We drop the traffic signs the numbers of which are less than 100 and select 23923 traffic signs (45 categories) of 10267 pictures from the Chinese traffic-sign benchmark \cite{Zhe_2016_CVPR}. We denote this dataset as $DS_1$. Moreover, we collect the small dataset of street view pictures from target cities. The small dataset consists of 162 no-left-turn signs, 127 no-right-turn signs, 108 no-u-turn signs, 106 no-parking signs and 149 negative samples, denoted as $DS_2$. $DS_1$ and $DS_2$ are used to train the traffic sign detection model. 

    \subsubsection{\textbf{Street View Pictures for Driving Behavior Context Augmentation}}
    After driver perspective modeling, 79,855 and 47,088 street view pictures are collected in Xiamen and Chengdu, respectively from Baidu Map \footnote{https://map.baidu.com}, and these street view pictures are input to the traffic sign detection model to get the traffic rule information. Specifically, the field of view is a significant parameter while collecting street view pictures. According to the basic design criteria for state highway from the New Zealand Transport Agency, when the driving speed is 0 km/h, the driver's horizontal angle of field is $180^{\circ}$. When the speed increases to 60 km/h, the angle decreases to $74^{\circ}$, and when the speed increases to 80 km/h and 100 km/h, the angle decreases to $60^{\circ}$ and $40^{\circ}$, respectively. Generally, the speed of the vehicle on the urban road is lower than 60 km/h. Thus, the field of view used in street view collection is $74^{\circ}$.

    % \begin{table}[h]
    %   \renewcommand{\arraystretch}{1.3}
    %   \caption{The Number of Instance Per Category in $DS_2$}
    %   \label{table:histogram}
    %   \centering{}
    %   \begin{tabularx}{.3\textwidth}{ @{} m{2.5cm} m{2.5cm} @{} } 
    %     \toprule{}
    %     \bfseries Category &\bfseries Number \\
    %     \hline
    %     No-left-turn &162 \\
    %     No-right-turn &127 \\
    %     No-u-turn &108 \\
    %     No-parking &106 \\
    %     Negative samples &149 \\
    %     \bottomrule
    %   \end{tabularx}
    % \end{table}

  \subsection{Experiment Settings}

    \subsubsection{Evaluation Plan}
    In traffic sign detection, we first separate $DS_1$ into training set and validation set, $80\%$ and $20\%$respectively, and train a traffic sign detection model based on the YOLOv3 \cite{redmon2018yolov3} model. Then from $DS_2$, we select $10\%$ of each type of traffic signs as the test set and we fine turn the model with the rest of the traffic signs, which is separated into training set and validation set, $80\%$ and $20\%$ respectively. The traffic sign detection model is used to detect traffic signs in 79,855 and 47,088 street view pictures collected from Xiamen and Chengdu, respectively. After that, we get the traffic violation distribution so as to infer the traffic violation-prone locations. In traffic violation-prone location inference, we show the thresholds of traffic violation-prone locations and the number of traffic violation-prone locations in Chengdu and Xiamen during a month, respectively. Finally we evaluate the runtime performance, present a traffic violation-prone location inference system, and give several case studies. 

    \subsubsection{Evaluation Metric}
    We evaluate the performance of the traffic sign detection model on the test set using the average precision of interested categories of traffic signs, i.e., $AP=\sum_{k=1}^Np(k)(r(k-1)-r(k))$, where $p(k)$ and $r(k)$ are the precision and recall at the $k$th threshold, and $mAP$ is the mean of $AP$ for all categories.

    \subsubsection{Baseline Methods}
    We compare our method with following traffic sign detection models.
    \begin{itemize}
      \item \textbf{AB \cite{bahlmann2005system}}: This method uses a set of Haar wavelet features obtained from AdaBoost training to detect traffic signs, and Bayesian generative modeling is used to classify traffic signs.
      \item \textbf{FC \cite{zhu2016traffic}}: This method consists of two deep learning components including fully convolutional network (FCN) guided traffic sign proposals and deep convolutional neural network (CNN) for object classification.
      \item \textbf{CNNs \cite{Zhe_2016_CVPR}}: This method uses a end-to-end convolutional neural network (CNN) to detect and classify traffic signs.
    \end{itemize}

  \subsection{Traffic Sign Detection Results}

    \begin{table}[!t]
      \renewcommand{\arraystretch}{1.3}
      \caption{The Evaluation Results on Traffic Sign Detection}
      \label{table:evaluation}
      \centering{}
      \begin{tabularx}{0.45\textwidth}{ @{} m{2cm} m{1cm} m{1cm} m{1cm} m{1cm} @{} } 
        \toprule{}
        \bfseries Signs &\bfseries AB & \bfseries FC &\bfseries CNNs &\bfseries Proposed \\
        \hline
        No-left-turn & 0.375 & 0.563 & 0.563 & \bfseries 0.688 \\
        No-right-turn& 0.462 & 0.693 & 0.538 & \bfseries 0.846 \\
        No-u-turn    & 0.727 & 0.455 & 0.545 & \bfseries 0.818 \\
        No-parking   & 0.182 & 0.636 & 0.818 & \bfseries 0.818 \\
        Total (mAP)  & 0.437 & 0.587 & 0.616 & \bfseries 0.793 \\
        \bottomrule
      \end{tabularx}
    \end{table}

    % Figure --------------------------------------
    \begin{figure}[!t]
    \vspace{-0.8cm}
      \centering
      \includegraphics[width=.47\textwidth]{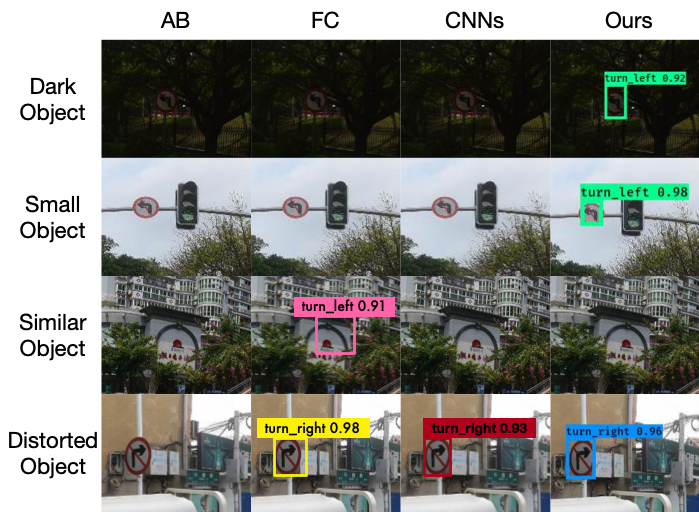}
      \caption{The performance of traffic sign detection models under severe conditions.}
      \label{fig:extreme_cases}
      \vspace{-0.5cm}
    \end{figure}

  The evaluation results on traffic sign detection are shown in TABLE. \ref{table:evaluation}. The street view pictures are taken in complicated urban environment with interferences and have low resolution. Besides, the size of traffic signs in the pictures are varied, and since the street view pictures are panoramic and the shooting angles are different, the edges of pictures are usually kind of distorted, which increases the difficulties of traffic sign detection. According to the evaluation results, the AB, FC, and CNNs methods do not perform well on our test set, and the fine-tuned YOLOv3 model in our proposed framework achieves the best performance. Furthermore, \figurename{~\ref{fig:extreme_cases}} shows examples of traffic sign detection under severe conditions. The AB method does not perform well when the traffic sign is in dark environment, very small, or distorted, but it rarely misidentify similar objects as traffic signs, while the FC method recognizes similar objects in pictures as traffic signs more frequently. The CNNs method works well when traffic signs are clear and it rarely misidentify similar object, but it fails when traffic signs are blurred. Different from the baseline methods, the proposed fine-tuned YOLOv3 model works well under these severe conditions.

    % \begin{table}[!h]
    %     \renewcommand{\arraystretch}{1.3}
    %     \caption{Summary of Traffic signs Detected in Xiamen and Chengdu}
    %     \label{table:traffic_sign_xiamen_chengdu}
    %     \centering{}
    %     \begin{tabularx}{0.4\textwidth}{ @{} m{2cm} m{2cm} m{2cm} @{} } 
    %       \toprule{}
    %       \bfseries Items &\bfseries Xiamen & \bfseries Chengdu \\
    %       \hline
    %       No-left-turn & 282 & 198 \\
    %       No-right-turn& 110 & 71 \\
    %       no-u-turn    & 29  & 19 \\
    %       No-parking   & 848 & 344\\
    %       Total        & 1269& 632\\
    %       \bottomrule
    %     \end{tabularx}
    %   \end{table}

      \begin{table}
        \renewcommand{\arraystretch}{1.3}
        \caption{Summary of Traffic signs Detected in Xiamen and Chengdu}
        \label{table:traffic_sign_xiamen_chengdu}
        \centering{}
        \begin{tabularx}{0.45\textwidth}{ @{} m{0.7cm} m{1cm} m{1.2cm} m{1cm} m{1cm} m{1cm} @{} } 
          \toprule{}
          \bfseries City &No-left-turn & No-right-turn & No-u-turn & No-parking & Total \\
          \hline
          Xiamen  & 282 & 110 & 29 & 848 &1269 \\
          Chengdu & 198 & 71 & 19 & 344 & 632 \\
          \bottomrule
        \end{tabularx}
        \vspace{-0.5cm}
      \end{table}
    
\begin{figure}[!t]
    \vspace{-0.1cm}
      \centering
      \includegraphics[width=.4\textwidth]{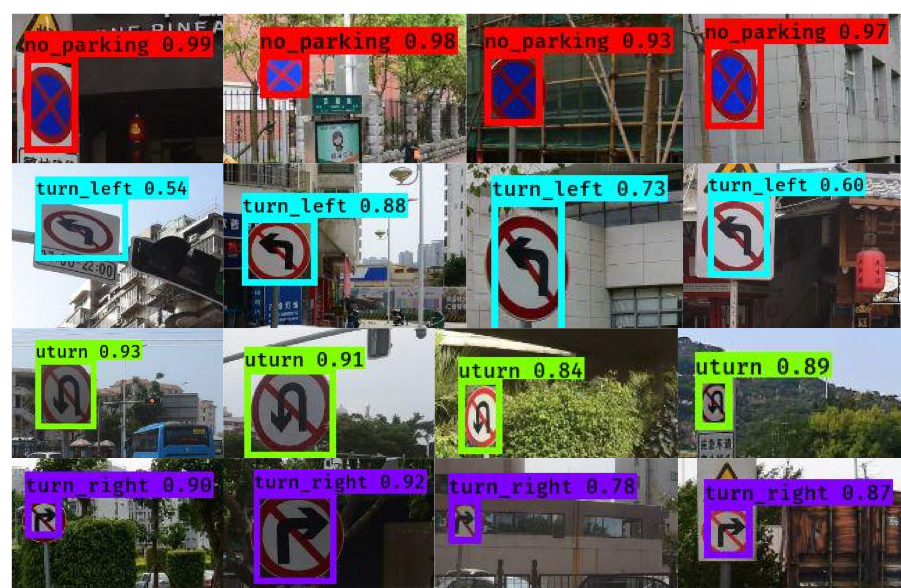}
      \caption{Some examples of traffic sign detected in street view pictures.}
      \label{fig:detection-results}
    \end{figure}

    Therefore, we use the fine-tuned YOLOv3-based traffic sign detection model to detect the traffic signs in 79,855 and 47,088 street view pictures collected from Xiamen and Chengdu, respectively. The summary of the traffic signs detected from the street view pictures in Xiamen and Chengdu are shown in TABLE. \ref{table:traffic_sign_xiamen_chengdu} and \figurename~\ref{fig:detection-results} shows some examples of traffic signs detected in street view pictures.

    % Figure --------------------------------------

    \subsection{Traffic Violation-Prone Location Inference Results}
\begin{figure}[!t]
      \vspace{-0.3cm}
      \centering
      \subfloat[Xiamen.]{
                \centering
                \includegraphics[width=.47\textwidth]{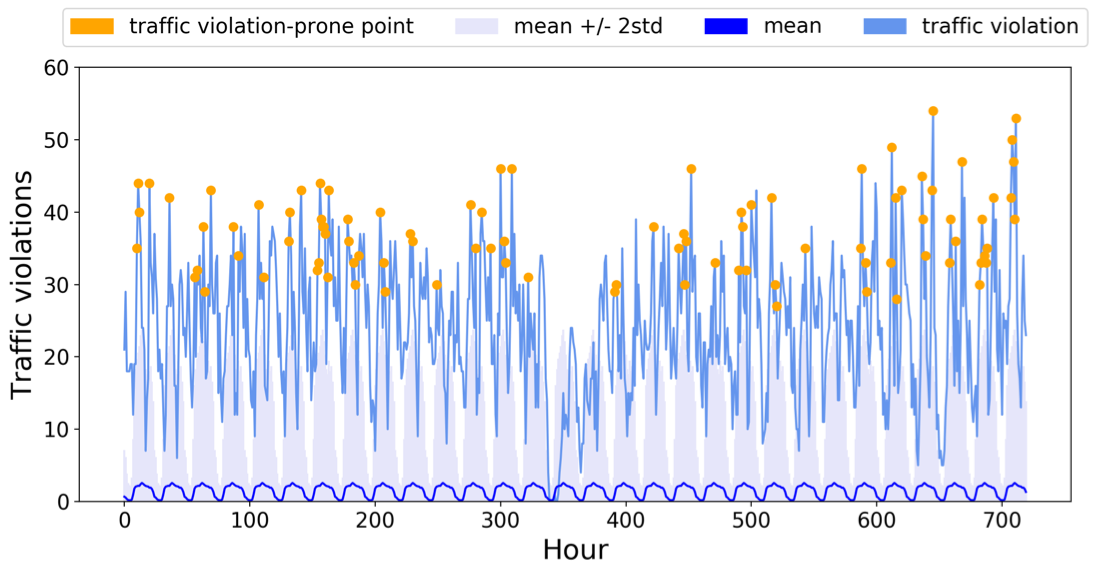}
        \label{fig:tvinfer-xm}
        }\\\vspace{-0.3cm}
        \subfloat[Chengdu.]{    
                \centering
                \includegraphics[width=.47\textwidth]{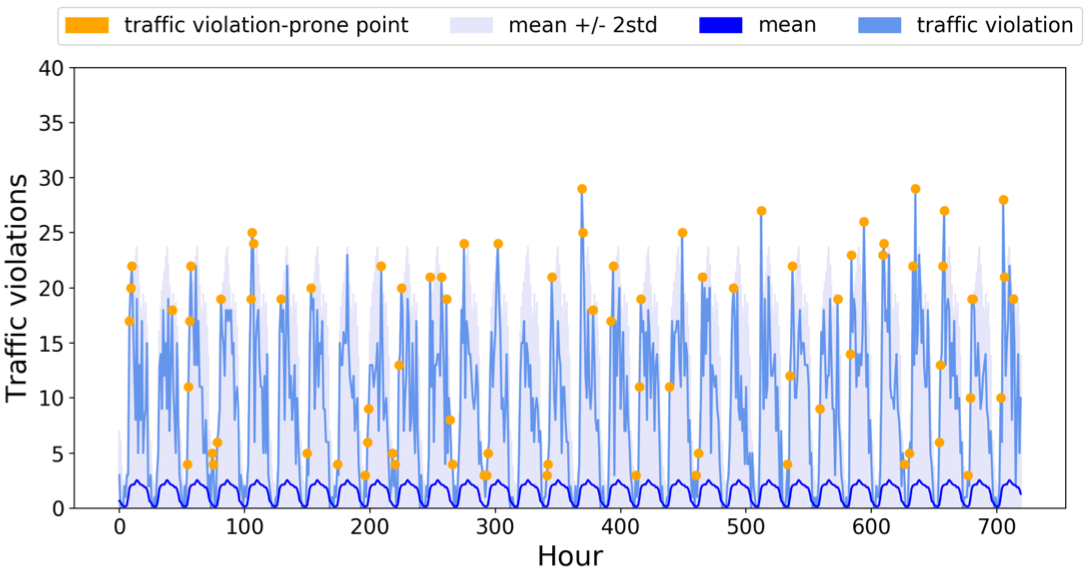}
            \label{fig:tvinfer-cd}
        }
        \caption{The thresholds of traffic violation-prone locations in Xiamen or Chengdu, and the points in each city represent the inferred traffic violation-prone points of a location in the city in a month. 'std' means standard deviation.}
      \label{fig:tvinfer}
    \end{figure}

\begin{figure}[!t]
    \vspace{-0.5cm}
      \centering
      \includegraphics[width=.5\textwidth]{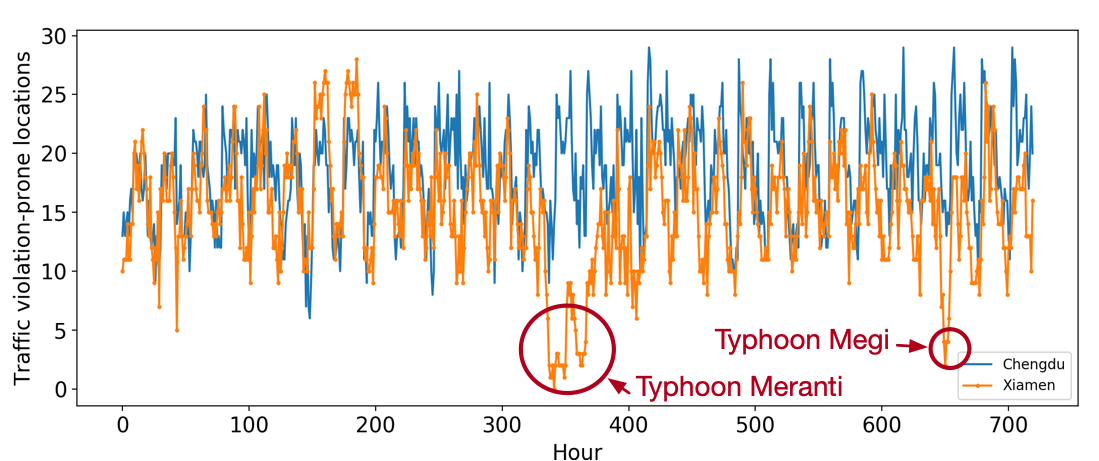}
      \caption{The number of traffic violation-prone locations in Chengdu (November) and Xiamen (September) in 2016.}
      \label{fig:tvloc_dis}
    \end{figure}

     \figurename~\ref{fig:tvinfer} shows the thresholds of traffic violation-prone locations in Xiamen and Chengdu, respectively, and the points marked in the figure represent the inferred traffic violation-prone points of a location in the corresponding city in a month. \figurename~\ref{fig:tvloc_dis} shows the number of traffic violation-prone locations in Chengdu, November, 2016, and Xiamen, September, 2016, respectively. Specifically, there were two typhoons influenced Xiamen in this month, Meranti on September 15th and Megi on September 27th, which are corresponding to the lowest two segments on the curve of Xiamen.

    % Figure --------------------------------------
    % Figure --------------------------------------

 \subsection{Runtime Performance}

    The typical temporal profiles of traffic violation-prone location candidates are updated every day when data from a new day have been collected. We deploy our system on a server with NVIDIA GeForce GTX 1080 Ti 11GB and 32GB RAM, and it takes an average of 55.937 ms and 0.209 ms to do traffic sign detection for one picture and map matching for one GPS point, respectively. In Xiamen, the average number of GPS points and street view pictures that need to be processed every day were about 11,743,359 and 2,662. In Chengdu, these numbers are 36,553,615 and 1,570, respectively. Therefore, the average time of processing new data collected every day is about 43.38 minutes in Xiamen and 128.79 minutes in Chengdu.

    % Figure --------------------------------------
    \begin{figure}[!t]
    \vspace{-0.3cm}
      \centering
      \includegraphics[width=.5\textwidth]{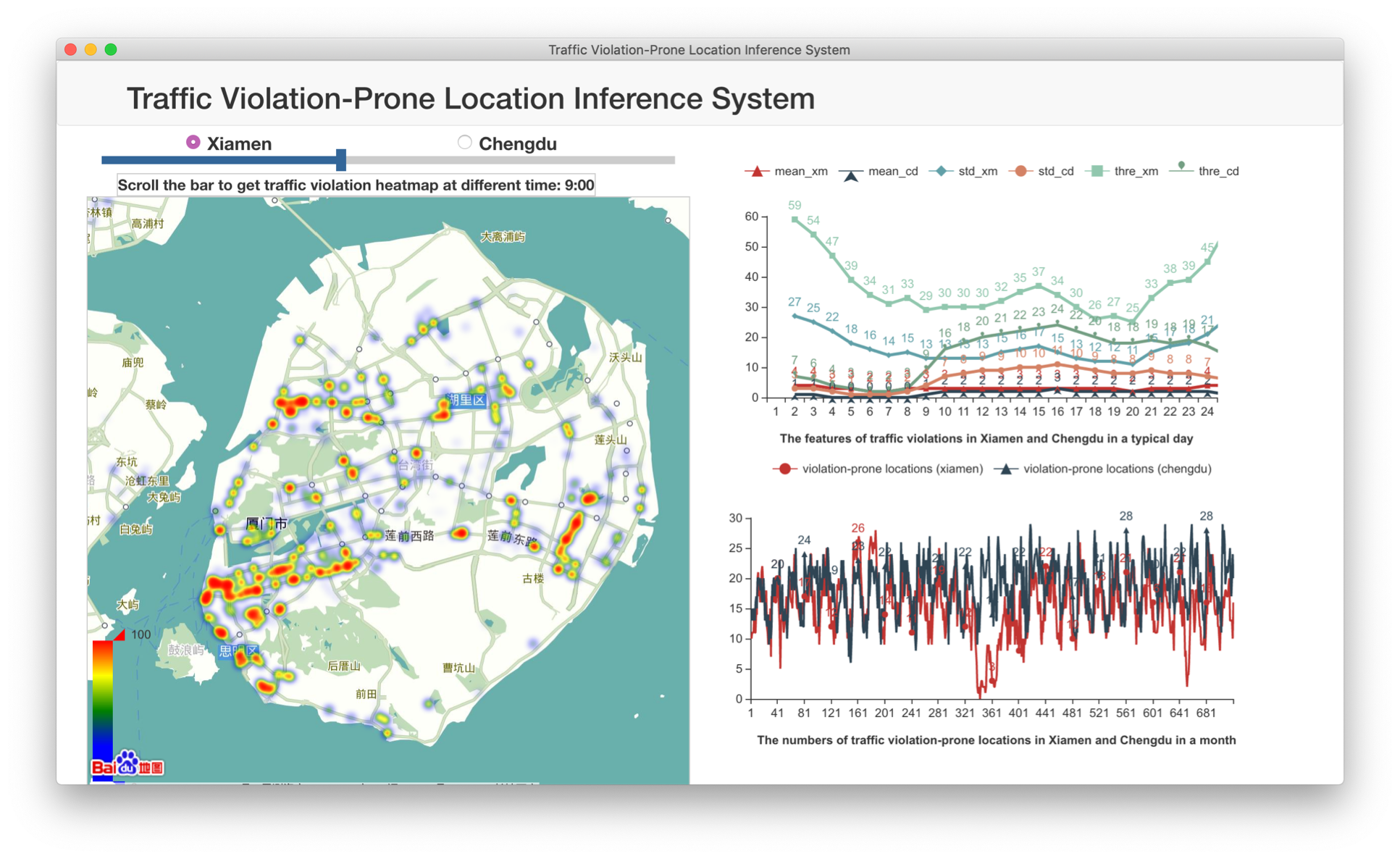}
      \caption{Traffic Violation-Prone Location Inference System.}
      \label{fig:system}
      \vspace{-0.3cm}
    \end{figure}

    \subsection{Traffic Violation-Prone Location Inference System}
    We build a traffic violation-prone location inference system \footnote{The system can be visited at: http://zhihanjiang.com/tv-infer-web/}. From this system, we can easily find the traffic violation-prone locations in a city at different time, as shown in \figurename~\ref{fig:system}.

    % Figure --------------------------------------
    \begin{figure}[!t]
      \centering
      \vspace{-0.3cm}
        \subfloat[]{
            \begin{minipage}[c]{.6\linewidth}
                \centering
                \includegraphics[width=\textwidth]{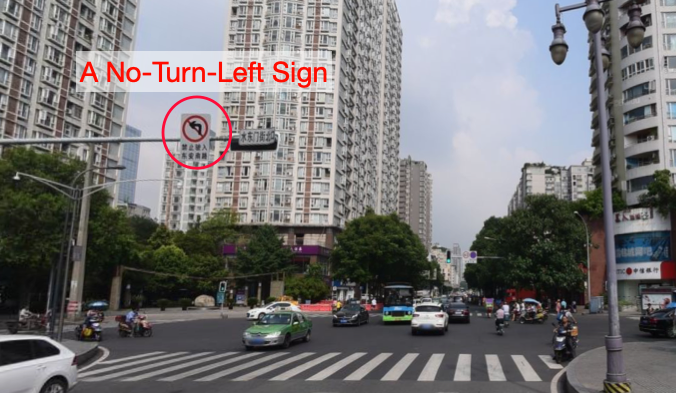}
            \end{minipage}
        \label{fig:case_study3_5}
        }
        \subfloat[]{    
            \begin{minipage}[c]{.35\linewidth}
                \centering
                \includegraphics[width=\textwidth]{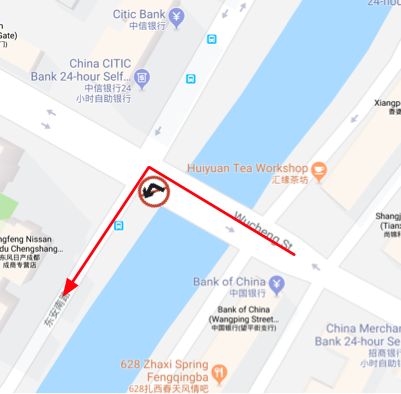}
            \end{minipage}
            \label{fig:case_study3_6}
        }
        \caption{An illegal turning-prone location in Chengdu. (a) The street view picture of an illegal turning-prone location. (b) An illegal turning-prone location}
      \label{fig:case_study5}
      % \vspace{-0.3cm}
    \end{figure}

    \subsection{Case Studies}

    \subsubsection{Illegal Turning} Illegal turning is the second most frequent traffic violation in Chengdu, and we found a lot of illegal turning behaviors happen on the intersection of Wucheng Street and Dongan South Road, as shown in \figurename~\ref{fig:case_study5}. In this intersection, drivers are forbidden to turn left (the red arrow in the figure) after crossing a bridge, but there are still many people violating the traffic regulation. 

    \subsubsection{Illegal Parking} 
\begin{figure}[!t]
      \vspace{-0.3cm}
      \centering
      \subfloat[]{
            \begin{minipage}[c]{.47\linewidth}
                \centering
                \includegraphics[width=\textwidth]{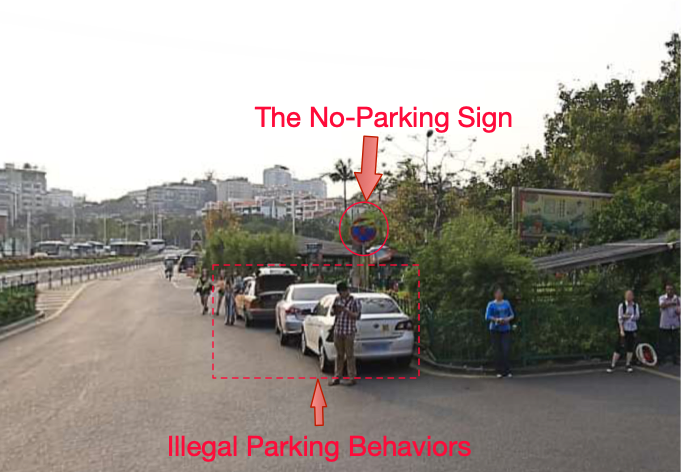}
            \end{minipage}
        \label{fig:case_study1_pk1}
        }
        \subfloat[]{    
            \begin{minipage}[c]{.47\linewidth}
                \centering
                \includegraphics[width=\textwidth]{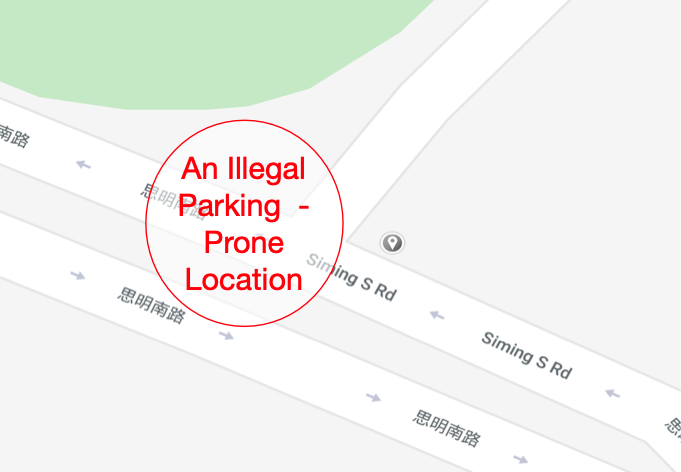}
            \end{minipage}
            \label{fig:case_study1_pk2}
        }
      \caption{An illegal parking-prone location in Xiamen. (a) The street view picture of an illegal parking-prone location in Xiamen. (b) An illegal parking-prone location.}
      \label{fig:case_study1_pk}

    \end{figure}

According to the Xiamen Traffic Police, illegal parking is the most frequent traffic violation from 2015 to 2018. Through observing heat maps, we can find that the intersection of Chenggong Avenue and Nanshan Road had a large number of illegal parking behaviors. However, after an in-depth investigation, we found that a new subway station was being built during that time, which led to terrible traffic conditions there. Thus there were many illegal parking behaviors. Besides, Xiahe Road, Siming South Road, and many other roads near the tourist hotspots are also traffic violation-prone locations. Siming South Road is a very busy road in Xiamen, along which there are a lot of tourist hotspots, such as Xiamen University, Nanputuo Temple, Overseas Chinese Museum, and it also intersects Zhongshan Road, which is a famous tourist road. The complicated environment and the large traffic volume make it become the road with maximum illegal parking behaviors in Xiamen. As shown in \figurename~\ref{fig:case_study1_pk}, some vehicles are parked on the road, although there is a no-parking sign erected there.

    % Figure --------------------------------------

    \subsubsection{Speeding} 
\begin{figure}[!t]
      \vspace{-0.8cm}
      \centering
      \subfloat[]{
            \begin{minipage}[c]{.62\linewidth}
                \centering
                \includegraphics[width=\textwidth]{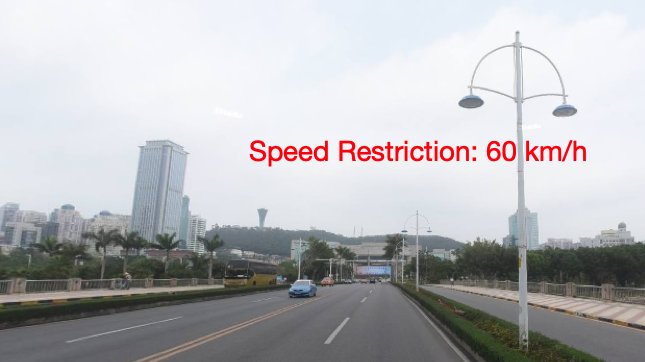}
            \end{minipage}
        \label{fig:case_study2_1}
        }
        \subfloat[]{    
            \begin{minipage}[c]{.32\linewidth}
                \centering
                \includegraphics[width=\textwidth]{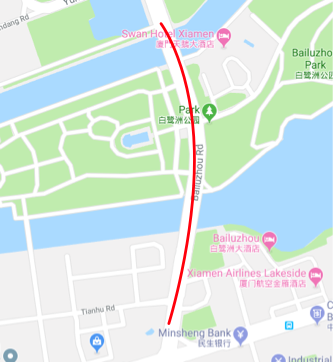}
            \end{minipage}
            \label{fig:case_study2_2}
        }
        \caption{A speeding-prone location in Xiamen. (a) The street view picture of a speeding-prone location.  (b) A speeding-prone location.}
      \label{fig:case_study2}
    \end{figure}

Speeding is the sixth most frequent traffic violation in Xiamen. \figurename~\ref{fig:case_study2} shows a speeding behavior on Bailuzhou Road in Xiamen. Bailuzhou Road is a speeding-prone location in Xiamen, which is a 4-lane dual carriageway. The speed restriction of Bailuzhou Road is 60 km/h. Besides we found a lot of speeding behaviors on Jiahe Road and Chenggong Avenue. They are both important roads crossing over Xiamen Island.
    % Figure --------------------------------------

	% In the future, we plan to model the traffic violation distribution and predict the traffic violation hotspots in cities, 

% ==========================================================
\section{Conclusion}

  In this work, we propose a low-cost, comprehensive and dynamic method for inferring the traffic violation-prone locations in cities based on the large-scale vehicle trajectory data and road environment data to provide some insights for the traffic management department about traffic dynamics in cities to help optimize the utility and effectiveness of the traffic enforcement strategies. Firstly, we normalize the trajectory data by map matching algorithms and get the driving behavior distribution. Secondly, we model match driving behaviors to corresponding road segments and restore the spatiotemporal contexts of driving behaviors to get the traffic rule information so that we can get three types of traffic violation distributions, i.e., illegal turning, illegal parking and speeding. Then we extract the spatiotemporal patterns of traffic violations to infer the traffic violation-prone locations in cities. Finally, we build a traffic violation-prone location inference system and give some case studies. The proposed method is evaluated using crowd-sensed, large-scale, and real-world datasets.

  In the future, we plan to broaden and deepen this work in two directions. Firstly, we plan to incorporate more trajectory open data sources from other cities. Secondly, we plan to incorporate urban environment data, such as Points of Interests and traffic volumes, to explore more in-depth relationships between traffic violation-prone locations and the urban environment.

\section*{Acknowledgment}
The authors would like to thank the reviewers for their constructive suggestions. This research is supported by NSF of China (61802325, 61872306 and U1605254), NSF of Fujian Province (2018J01105), the China Fundamental Research Funds for the Central Universities (20720170040), and Xiamen Science and Technology Bureau (3502Z20193017).
\ifCLASSOPTIONcaptionsoff
  \newpage
\fi

% \bibliographystyle{IEEEtran}
% \bibliography{system2020}



% \end{thebibliography}

% % biography section
% % 
% % If you have an EPS/PDF photo (graphicx package needed) extra braces are
% % needed around the contents of the optional argument to biography to prevent
% % the LaTeX parser from getting confused when it sees the complicated
% % \includegraphics command within an optional argument. (You could create
% % your own custom macro containing the \includegraphics command to make things
% % simpler here.)


% % insert where needed to balance the two columns on the last page with
% %\newpage


% % You can push biographies down or up by placing
% % a \vfill before or after them. The appropriate
% % use of \vfill depends on what kind of text is
% % on the last page and whether or not the columns
% % are being equalized.

% %\vfill

% % Can be used to pull up biographies so that the bottom of the last one
% % is flush with the other column.
% %\enlargethispage{-5in}

% % that's all folks
\end{document}